\documentclass[journal,draftclsnofoot,onecolumn,12pt]{IEEEtran}
\usepackage{amsmath}
\usepackage{amssymb}
\usepackage{amsfonts}
\usepackage{color}
\usepackage[noadjust]{cite}
\usepackage{graphicx} 
\usepackage{subfigure}
\usepackage[linesnumbered,ruled]{algorithm2e}

\newcommand{\argmax}[1]{\underset{#1}{\operatorname{arg}\,\operatorname{max}}\;}

\ifodd 1
\usepackage{soul}
\usepackage{color}
\setstcolor{red}

\newcommand{\del}[1]{\st{#1}} 

\newcommand{\com}[1]{\textbf{\color{red} (COMMENT: #1)}} 
\newcommand{\response}[1]{\textbf{\color{green} (RESPONSE: #1)}} 
\else

\newcommand{\del}[1]{}

\newcommand{\com}[1]{}
\newcommand{\comg}[1]{}
\newcommand{\response}[1]{}
\fi



\begin{document}
\title{{Intelligent Reflecting Surface Meets OFDM: Protocol Design and Rate Maximization}
\thanks{Part of this work will be presented at the IEEE Global Communications Conference (Globecom), Waikoloa, HI, USA, 2019 \cite{irsglobecom}.}
\thanks{Y. Yang is with the Department of Electrical and Computer Engineering, National University of Singapore, and also with Halliburton Far East, Singapore (e-mail: yifeiyang@u.nus.edu).}
\thanks{B. Zheng, S. Zhang, and R. Zhang are with the Department of Electrical and Computer Engineering, National University of Singapore (e-mail: \{elezbe, elezhsh, elezhang\}@nus.edu.sg). (Corresponding author: S. Zhang).}
}

\author{\IEEEauthorblockN{Yifei~Yang, Beixiong~Zheng,~\IEEEmembership{Member,~IEEE}, Shuowen~Zhang,~\IEEEmembership{Member,~IEEE}, and Rui~Zhang,~\IEEEmembership{Fellow,~IEEE}}}

\maketitle

\begin{abstract}
Intelligent reflecting surface (IRS) is a promising new technology for achieving both spectrum and energy efficient wireless communication systems in the future. 
However, existing works on IRS mainly consider frequency-flat channels and assume perfect knowledge of channel state information (CSI) at the transmitter. Motivated by this, in this paper we study an IRS-enhanced orthogonal frequency division multiplexing (OFDM) system under frequency-selective channels and propose a practical transmission protocol with channel estimation. First, to reduce the overhead in channel training and estimation and to exploit the channel spatial correlation, we propose a novel IRS elements grouping method, where each group consists of a set of adjacent IRS elements that share a common reflection coefficient. Based on this grouping method, we propose a practical transmission protocol where only the combined channel of each group needs to be estimated, thus substantially reducing the training overhead. Next, with any given grouping and estimated CSI, we formulate the problem to maximize the achievable rate by jointly optimizing the transmit power allocation and the IRS passive array reflection coefficients. Although the formulated problem is non-convex and thus difficult to solve, we propose an efficient algorithm to obtain a high-quality suboptimal solution for it, by alternately optimizing the power allocation and the passive array coefficients in an iterative manner, along with a customized method for the initialization. Simulation results show that the proposed design significantly improves the OFDM link rate performance as compared to the case without using IRS. Moreover, it is shown that there exists an optimal size for IRS elements grouping which achieves the maximum achievable rate due to the trade-off between the training overhead and IRS passive beamforming flexibility. 
\end{abstract}

\section{Introduction}
The explosion of mobile data and the ever-increasing demand for higher data rates have continuously driven the advancement of wireless communication technologies in the past decade, such as polar code, massive multiple-input multiple-output (MIMO) and millimeter wave (mmWave) communications, among others. Moreover, a 1000-fold increment in network capacity with ubiquitous connectivity and low latency is envisioned for the forthcoming fifth-generation (5G) wireless network \cite{5g}. Meanwhile, the energy efficiency of future wireless networks is expected to be improved by several orders of magnitude so as to maintain the power consumption at increasingly higher data rates. Although massive MIMO and mmWave, seen as the key enablers for 5G, are expected to achieve dramatic spectral efficiency improvements, the deployment of large-scale antenna arrays usually results in high implementation cost and increased power consumption \cite{mimoover}. In addition, combining mmWave with massive MIMO for further performance improvements generally requires more sophisticated signal processing as well as more costly and energy-consuming hardware. Hence, finding green and sustainable solutions to enhance wireless network performance for higher data rates and efficiency still remains crucial.  

Recently, intelligent reflecting surface (IRS) or its various equivalents have been proposed as a promising new solution to achieve the above goals \cite{qqmag, qqtwc,qqicassp, debtwc, mmwre,debtwc2,irsstatscsi,renzomag}. Specifically, IRS is a reconfigurable planar array comprising a vast number of passive reflecting elements, each of which is able to independently induce a phase shift and/or an attenuation to the incident signal and thereby collaboratively alter the reflected signal propagation to achieve desired channel responses in wireless communications. By properly adjusting the phase shifts and attenuations of the IRS's elements, their reflected signals can be combined with those from other paths constructively at the receiver to enhance the received signal power and/or destructively to suppress the co-channel interference, both leading to improved wireless link performance \cite{qqmag}. Hence, different from the conventional half-duplex amplify-and-forward (AF) relay, IRS achieves high beamforming gains by intelligent reflection in a \emph{full-duplex} manner, without consuming any energy or requiring additional time/frequency resource for signal re-generation and re-transmission. Although passive reflect-array antennas have been widely applied in radar and satellite communication systems, their applications in mobile wireless communications are rather limited. This is because traditional passive arrays only allow for fixed phase-shift patterns once fabricated and are thus unable to adapt to the dynamic wireless channel induced by user mobility. Fortunately, the advances in radio frequency (RF) micro electromechanical systems (MEMS) and metamaterial (e.g., metasurface) have made it feasible to reconfigure the phase shifts in real time \cite{metasur}, thus greatly enhancing the functionality and applicability of IRS for wireless communications.

Existing works on IRS \cite{qqtwc,qqicassp, debtwc,mmwre,irssw} mainly assume perfect channel state information (CSI) at the base station (BS)/IRS, based on which the design parameters are optimized to enhance the system performance. In \cite{debtwc2} and \cite{irsstatscsi}, only statistical CSI is required for the design problem since large-scale antenna array systems are deployed and therefore deterministic equivalent results by exploiting the large system limit can be obtained. On the other hand, for IRS-aided wireless systems where knowledge of instantaneous CSI is required, there are in general two approaches for the channel acquisition, depending on whether the IRS elements are equipped with receiving RF chains \cite{qqmag}. For the first approach where IRS elements are capable of both sensing and reflecting, conventional channel estimation methods can be readily applied. Specifically, channels between the IRS and the BS/users can be estimated by leveraging channel reciprocity and time division duplexing (TDD). However, equipping the IRS elements with (receive) RF chains generally results in higher implementation cost and energy consumption. In contrast, for the second approach where no sensors are installed on the IRS (i.e., the IRS elements are purely passive), it is impossible to explicitly obtain the individual channels of the BS-IRS and IRS-user links. One possible method for this (more challenging) case presented in \cite{qqmag} is to bypass the channel estimation and design the passive beamforming coefficients directly based on the feedback from the BSs/users (e.g., using codebook-based passive beamforming), which, unfortunately, can be time-consuming due to practically vast number of IRS elements and the potentially large codebook size required to achieve reasonably good beamforming performance. In general, there has been very limited work on how to design an efficient channel estimation strategy specifically catered to the IRS-based passive array systems. Moreover, it is worth noting that prior works on IRS-aided wireless systems \cite{qqtwc,qqicassp, debtwc,debtwc2,mmwre,irsstatscsi} have mainly considered frequency-flat (non-selective) fading channels for narrowband communications, where the IRS reflection coefficients are designed to align the phase of the BS-IRS-user reflected link with that of the BS-user direct link for constructive superposition. However, when frequency-selective channels are considered, the IRS reflection coefficients need to cater to all signal paths at different delays, for which the underlying optimization problem is thus more challenging to solve. Notwithstanding, the design of IRS passive array coefficients over frequency-selective fading channels for broadband communications has not been addressed yet, to the authors' best knowledge.

\begin{figure}[t]
    \centering
    \includegraphics[width=0.6\linewidth, keepaspectratio]{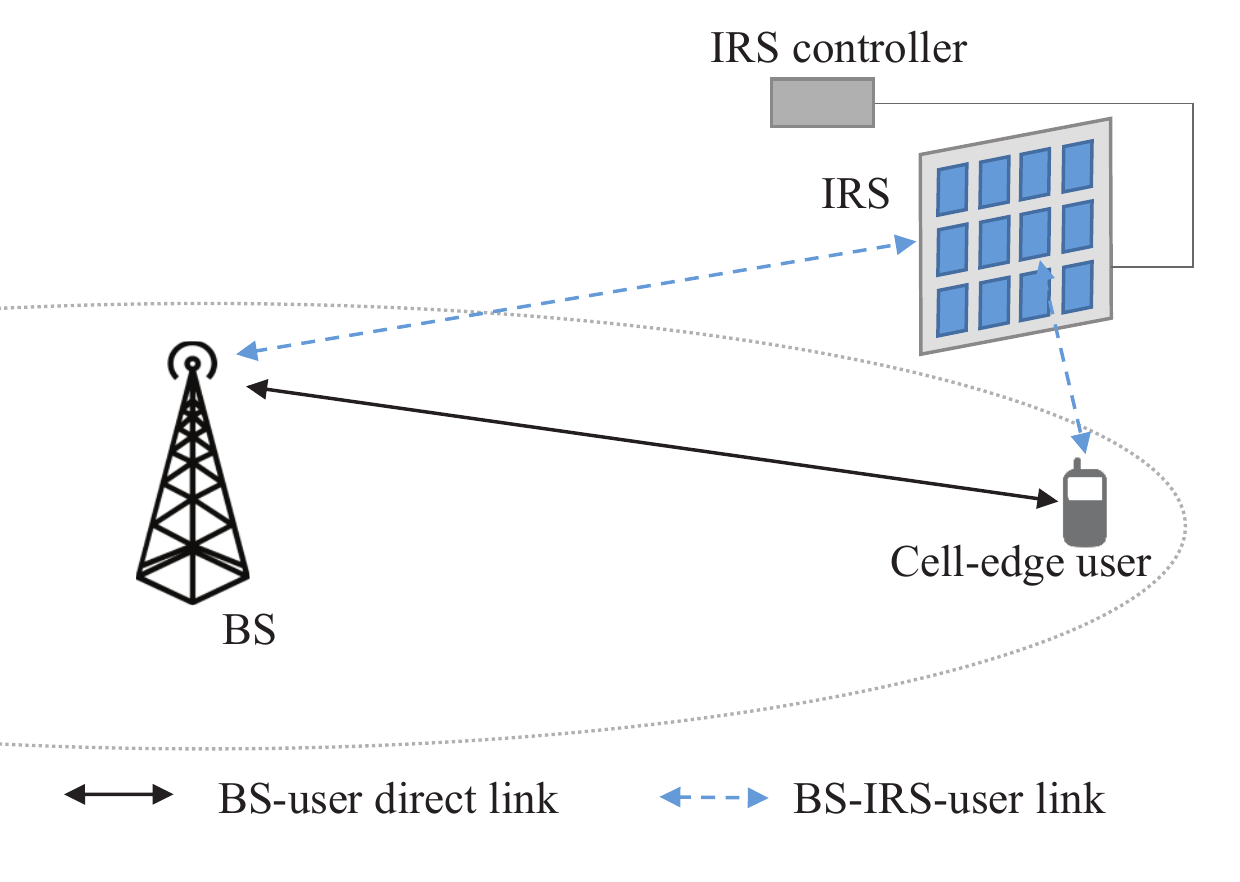}
    \DeclareGraphicsExtensions.
    \caption{An IRS-enhanced wireless system.}
    \label{fig:irs}
    \vspace{-6mm}
\end{figure}

Motivated by the above challenges, in this paper, we investigate an IRS-aided orthogonal frequency division multiplexing (OFDM)-based wireless system over frequency-selective channels, as shown in Fig. \ref{fig:irs}. We consider point-to-point communication between a BS and a single user in the vicinity of an IRS, where the IRS and the user are both far away from the BS (e.g., in a cell-edge user scenario). Our considered scenario corresponds to practical multiuser wireless systems based on either time division multiple access (TDMA) or orthogonal frequency division multiple access (OFDMA) with a given set of OFDM subcarriers (SCs) assigned to the user of interest. As the IRS is usually equipped with a large number of reflecting elements, how to jointly design their reflection coefficients (i.e., phase shifts and amplitude attenuations) so as to achieve the optimal constructive superposition of the signals reflected by the IRS and those from other paths at the receiver is crucial to maximizing the link achievable rate. However, this is a non-trivial problem to solve under our considered setup due to the following reasons. First, the BS-IRS-user channels for all the IRS elements need to be estimated for designing their reflection coefficients, for which the required channel training overhead is proportional to the number of IRS reflecting elements, thus may be prohibitive or even unaffordable in practice. Second, under the frequency-selective channel with multiple paths, the reflection coefficients of the IRS need to cater to the channel gains and delays of all paths to the receiver, including both the reflected paths by the IRS and the remaining non-reflected paths directly from the transmitter. Finally, the achievable rate for the user is determined by both the IRS reflection coefficients and the transmit power allocation over OFDM SCs, which are intricately coupled and thus need to be jointly optimized. In general, the size of the joint optimization problem increases with the number of IRS elements and/or OFDM SCs. In this paper, we tackle the aforementioned challenges via novel protocol and algorithm designs. The main contributions of this paper are summarized as follows:
\begin{itemize}
	\item First, we propose a novel design by dividing the IRS elements into groups, where each group consists of adjacent elements that are assumed to share a common reflection coefficient. Hence, only the combined channel of each group needs to be estimated. Based on this grouping method, a practical transmission protocol with pilot training is further proposed, where the combined channel associated with each group is estimated sequentially by turning on only the IRS elements of the corresponding IRS group while the other elements are set to be off. Note that compared to the straightforward approach of estimating the channel of each of the IRS's elements and jointly designing their reflection coefficients, grouping of the IRS elements exploits the channel correlation among adjacent IRS elements and significantly reduces the required training overhead as well as the IRS coefficient design complexity. It is also worth noting that the proposed channel estimation does not require any receive RF chains or sensors to be installed on the IRS, thus featuring low implementation cost and energy consumption.
	\item Next, based on the proposed IRS elements grouping and transmission protocol, we formulate a new optimization problem aiming to maximize the achievable rate by jointly optimizing the power allocation at the transmitter and the passive array coefficients at the IRS with the estimated channels. However, the formulated problem is non-convex and thus difficult to solve optimally. This motivates us to propose an iterative algorithm that alternately optimizes the power allocation and passive array coefficients, which is guaranteed to converge to at least a locally optimal solution. Moreover, we devise a customized initialization method for the iterative algorithm, by considering the IRS coefficient design that maximizes the overall channel power between the BS and the user.
	\item Finally, we evaluate the performance of our proposed designs via numerical results. It is shown that the proposed joint power allocation and IRS coefficient designs generally achieve improved rate performance compared to systems without using IRS or with random IRS reflection coefficients, for both cases of perfect and estimated CSI. Moreover, it is shown that the proposed initialization scheme is already able to achieve comparable performance to that of the iterative algorithm, thus is suitable for low-complexity implementation in practice. Furthermore, by comparing the performance of the proposed grouping with different group size and at high 
		or low 
		signal-to-noise ratio (SNR) values, it is unveiled that there generally exists an optimal group size that maximizes the achievable rate considerably, with a given channel coherence time, by optimally striking the balance between minimizing the training overhead and maximizing the IRS passive beamforming flexibility. 
\end{itemize}

The rest of this paper is organized as follows. Section \ref{sec:sysmodel} presents the system model of the IRS-enhanced OFDM communication and illustrates the proposed IRS elements grouping. Section \ref{sec:ce} proposes a practical transmission protocol with channel estimation based on the proposed grouping scheme. Given grouping and channel estimates, Section \ref{sec:pfsol} formulates the joint optimization problem of the OFDM power allocation and IRS reflection coefficients to maximize the achievable rate, and proposes an iterative algorithm to obtain a suboptimal solution for it. Section \ref{sec:sim} presents numerical results to evaluate the performance of the proposed protocol and designs. Finally, Section \ref{sec:con} concludes the paper. 

\emph{Notation:} For ease of reference, Table \ref{table:not} summarizes the main variables which will be used throughout this paper. Furthermore, vectors and matrices are denoted by bold-face lower-case and upper-case letters, respectively. Sets are designated by upper-case calligraphic letters. $[\boldsymbol{x}]_{a:b}$ denotes the subvector that contains the $a$-th to $b$-th elements of $\boldsymbol{x}$. $\Re({\cdot})$, $\Im({\cdot})$, $(\cdot)^*$ and $\arg(\cdot)$ denote the real part, imaginary part, conjugate, and angle of a complex number, respectively. For a matrix $\boldsymbol{M}$ of arbitrary size, $\boldsymbol{M}^H$ denotes the conjugate transpose, and $\boldsymbol{M}_{i,j}$ denotes the entry in the $i$-th row and $j$-th column. $\mathrm{Tr}(\cdot)$ denotes the matrix trace. $\boldsymbol{F}_N$ and $\boldsymbol{F}_N^H$ refer to the $N\times N$ discrete Fourier transform (DFT) and inverse DFT (IDFT) matrices, respectively. $\boldsymbol{0}_{a \times b}$ denotes an all-zero matrix of size $a\times b$, $\boldsymbol{1}_{a \times b}$ denotes an all-one matrix of size $a\times b$, $\boldsymbol{I}_M$ denotes the identity matrix of size $M\times M$, and $\boldsymbol{e}_m$ denotes the $m$-th column of $\boldsymbol{I}_M$. $\mathrm{diag}(\boldsymbol{x})$ denotes a square diagonal matrix with the elements of $\boldsymbol{x}$ on the main diagonal, whereas $\mathrm{Diag}(\boldsymbol{X})$ denotes a column vector formed by the main diagonals of $\boldsymbol{X}$. $\|\cdot\|$ denotes the $l_2$ norm and $\left\|\cdot\right\|_F$ denotes the Frobenius norm. $\otimes$ denotes the Kronecker product and $*$ denotes the linear convolution. The distribution of a circularly symmetric complex Gaussian (CSCG) random variable with mean $\mu$ and variance $\sigma^2$ is denoted by $\mathcal{CN}(\mu,\sigma^2)$; and $\sim$ stands for ``distributed as". $\mathbb{C}^{x\times y}$ denotes the space of $x\times y$ complex matrices. $\mathbb{R}$ denotes the space of real numbers. $\max(x,y)$ denotes the maximum between two real numbers $x$ and $y$. ${\mathbb E}\left\{ \cdot \right\}$ stands for the expectation operation and $\mathcal{O}(\cdot)$ stands for the standard big-O notation.

\begin{table}[t]
\centering
\caption{List of Main Variables and Their Physical Meanings}
\begin{tabular}{ | l | l |}   
    \hline
    $\boldsymbol{h}_d$ & BS-user (direct) channel \\ 
    $\boldsymbol{h}_r$ & BS-IRS-user (reflected) channel \\
    $\tilde{\boldsymbol{h}}$ & Overall impulse response for combined BS-user and BS-IRS-user channels\\
    $\boldsymbol{v}$ & Overall frequency response for combined BS-user and BS-IRS-user channels\\
    $\boldsymbol{h}_m$ & BS-IRS channel for the $m$-th IRS element \\
    $\boldsymbol{g}_m$ & IRS-user channel for the $m$-th IRS element \\
    $\boldsymbol{\nu}_m$ & BS-IRS and IRS-user composite channel for the $m$-th IRS element\\
    $\boldsymbol{\nu}'_k$ & BS-IRS and IRS-user composite channel for the $k$-th IRS group\\
    $\boldsymbol{V}$ & BS-IRS and IRS-user composite channel for all IRS elements \\
    $\boldsymbol{V}'$ & BS-IRS and IRS-user composite channel for all IRS groups \\
    $\boldsymbol{\phi}$ & IRS reflection coefficients\\
    $\bar{\boldsymbol{\phi}}$ & IRS group reflection coefficients\\
    \hline
    \end{tabular}
\label{table:not}
\end{table}

\section{System Model and IRS Elements Grouping Design}
\label{sec:sysmodel}
We consider a single-user OFDM-based wireless system, wherein an IRS is employed to enhance the communication between a BS and a user, as illustrated in Fig. \ref{fig:irs}. For the purpose of exposition, we assume that the BS and the user are both equipped with a single antenna. The IRS is assumed to comprise $M$ passive reflecting units, denoted by the set $\mathcal{M}=\{1,\dotsc, M\}$, and is connected to a controller, which adjusts the IRS pattern for desired signal reflection. Without loss of generality, we focus on uplink communication from the user to the BS in this paper, where pilot symbols are sent from the user and reflected by the IRS, based on which the BS estimates all the involved channels, computes the optimal design parameters, and then informs the IRS via a separate wireless control link (details to be given in Section \ref{sec:ce}). For downlink communication, by assuming channel reciprocity and leveraging TDD, the design parameters can be similarly optimized at the BS based on the channel information obtained during the uplink training. It is further assumed that the signals that are reflected by the IRS more than once have negligible power due to severe path loss and thus are ignored. We consider a quasi-static block fading channel model for all channels involved and focus on one particular fading block where the channels remain approximately constant. 

Similar to conventional OFDM-based systems, the total bandwidth of the system is equally divided into $N$ orthogonal SCs, denoted by the set $\mathcal{N}=\{0,\dotsc,N-1\}$. Moreover, let $\boldsymbol{p}=[p_0,\dotsc,p_{N-1}]^T\in\mathbb{R}^{N\times 1}$, where each $p_n\geq 0$ denotes the power allocated to the $n$-th SC at the BS. Assume the total transmission power available at the BS is $P$. Thus, the power allocation should satisfy $\sum_{n=0}^{N-1} p_n\leq P$. Let $\boldsymbol{h}_{d}=[\bar{h}_{0},\dotsc,\bar{h}_{L-1},\boldsymbol{0}_{1\times\left(N-L\right)}]^T\in\mathbb{C}^{N\times 1}$ denote the zero-padded $L$-tap baseband equivalent multipath channel for the BS-user direct link. Moreover, there exists an $L_0$-tap baseband equivalent multipath channel for the BS-IRS-user link, through which the signals transmitted by the user are reflected by the IRS before arriving at the BS. Specifically, let $\boldsymbol{h}_m\in\mathbb{C}^{L_1\times 1}$ denote the $L_1$-tap baseband equivalent BS-IRS channel for the $m$-th reflecting element at the IRS, $m\in\mathcal{M}$. Similarly, let $\boldsymbol{g}_m\in\mathbb{C}^{L_2\times 1}$ denote the $L_2$-tap baseband equivalent channel of the IRS-user link for the $m$-th reflecting element, $m\in\mathcal{M}$. At the IRS, each element re-scatters the received signals with an independent reflection coefficient. Specifically, let $\boldsymbol{\phi}=[\phi_1, \dotsc, \phi_M]^T\in\mathbb{C}^{M\times 1}$ denote the IRS reflection coefficients, where each $\phi_m=\beta_m e^{j\theta_m}$ comprises an amplitude coefficient $\beta_m\in[0,1]$ and a phase shift $\theta_m\in[-\pi,\pi)$, i.e., $|\phi_m|\leq 1$. The composite BS-IRS-user channel for the $m$-th reflecting element is thus the concatenation of the BS-IRS channel, IRS reflection, and IRS-user channel, which is given by $\boldsymbol{h}_m*\phi_m*\boldsymbol{g}_m=\phi_m\boldsymbol{h}_m*\boldsymbol{g}_m\in\mathbb{C}^{L_0\times 1}$, where $L_0=L_1+L_2-1$. For ease of exposition, define $\boldsymbol{V}=[\boldsymbol{\nu}_1,\dotsc,\boldsymbol{\nu}_M]\in\mathbb{C}^{N\times M}$ as the zero-padded concatenated BS-IRS and IRS-user channels, where each $\boldsymbol{\nu}_m\triangleq[(\boldsymbol{h}_m*\boldsymbol{g}_m)^T,\boldsymbol{0}_{1\times(N-L_0)}]^T\in\mathbb{C}^{N\times 1}$. The composite BS-IRS-user channel for all IRS reflecting elements, denoted by $\boldsymbol{h}_r\in\mathbb{C}^{N\times 1}$, can thus be expressed as 
\begin{equation}\label{eqn:hr1}
	\boldsymbol{h}_r=\boldsymbol{V}\boldsymbol{\phi}.
\end{equation}
Hence, the superposed channel impulse response (CIR) between the BS and the user by combining the BS-user (direct) channel and the BS-IRS-user (IRS-reflected) channel is given by
\begin{equation}\label{eqn:hall}
	\tilde{\boldsymbol{h}}=\boldsymbol{h}_{d}+\boldsymbol{h}_{r}.
\end{equation}

Assume OFDM modulation at the BS with a cyclic prefix (CP) of length $N_{\mathrm{CP}}$, with $N_{\mathrm{CP}}\geq\max(L, L_0)$. The channel frequency response (CFR) $\boldsymbol{v}=[v_{0},\dotsc,v_{N-1}]^T\in\mathbb{C}^{N\times 1}$ of the CIR $\tilde{\boldsymbol{h}}$ is thus given by
\begin{align}
	\boldsymbol{v}=\boldsymbol{F}_N\tilde{\boldsymbol{h}}.\label{eqn:vall}
\end{align}
It is worth noting that with given $\boldsymbol{\phi}$, knowledge of $\boldsymbol{\nu}_m$ suffices to characterize the BS-IRS-user composite channel for the $m$-th reflecting element, thus explicit knowledge of the individual channels $\boldsymbol{g}_m$ and $\boldsymbol{h}_m$ are not required. Hence, the overall CFR can be rewritten as $\boldsymbol{v}=\boldsymbol{F}_N\tilde{\boldsymbol{h}}=\boldsymbol{F}_N\left(\boldsymbol{h}_{d}+\boldsymbol{V}\boldsymbol{\phi}\right)$, and the CFR at each $n$-th SC is given by
\begin{equation}\label{eqn:vn}
	v_n=\boldsymbol{f}_n^H\boldsymbol{h}_{d}+\boldsymbol{f}_n^H\boldsymbol{V}\boldsymbol{\phi}, \quad n\in\mathcal{N},
\end{equation}
where $\boldsymbol{f}_n^H$ denotes the $n$-th row of the DFT matrix $\boldsymbol{F}_N$. Therefore, the maximum achievable rate in bits per second per Hertz (bps/Hz) is given by
\begin{align}
	r(\boldsymbol{p},\boldsymbol{\phi})\!=\!\frac{1}{N\!+\!N_{\mathrm{CP}}}\!\sum_{n=0}^{N-1}\! \log_2\!\left(\!1\!+\!\frac{|\boldsymbol{f}_n^H\boldsymbol{h}_{d}\!+\!\boldsymbol{f}_n^H\boldsymbol{V}\boldsymbol{\phi}|^2p_n}{\Gamma\sigma^2} \!\right)\!,\label{eqn:rk}
\end{align}
where $\Gamma\geq 1$ is the gap from channel capacity owing to a practical modulation and coding scheme (MCS); the receiver noise at each SC is modelled as an independent CSCG random variable with mean zero and variance $\sigma^2$. 

Note that (\ref{eqn:rk}) represents the theoretical upper-bound of the achievable rate of the considered OFDM system, which is difficult to be achieved in practice. This is because to perform OFDM coherent detection at the receiver as well as to design the transmit power allocation and IRS reflection coefficients, accurate knowledge of the CSI (i.e., $\boldsymbol{h}_d$ and $\boldsymbol{V}$) is required, which needs to be acquired at the cost of channel training and feedback overhead. Moreover, note that the dimension of the composite channel $\boldsymbol{V}$ grows linearly with the number of IRS reflecting elements, $M$, which can be very large in practice (e.g., current metasurfaces are typically equipped with more than tens and up to thousands of elements \cite{metasurcell}). The number of involved channel coefficients therein is therefore much larger than that for conventional OFDM systems without the IRS, which leads to increased overhead and higher complexity for channel training and estimation to obtain all of them. 

\begin{figure}[t]
    \centering
    \includegraphics[width=0.5\linewidth, keepaspectratio]{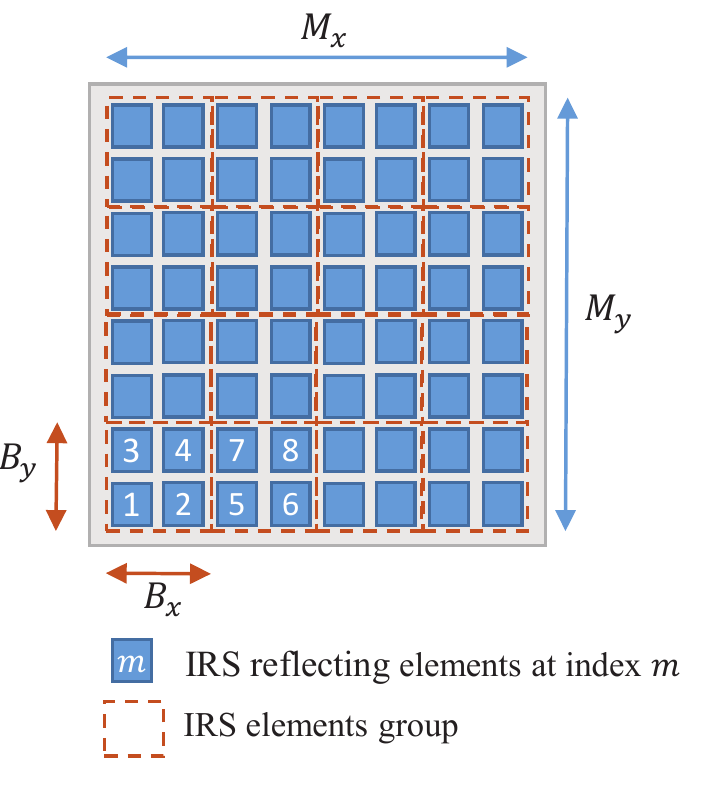}
    \DeclareGraphicsExtensions.
    \caption{Illustration of the IRS reflect array and proposed elements grouping.}
    \label{fig:irsgroup}
    \vspace{-6mm}
\end{figure}

To address the aforementioned issues, in this paper, we propose an \emph{IRS elements grouping} method to reduce the training overhead and estimation complexity. Specifically, note that since the IRS elements are usually tightly packed \cite{metasurcell}, the channels for \emph{adjacent} elements are practically \emph{correlated}. Therefore, we propose to group the adjacent IRS elements that form a small block, as illustrated in Fig. \ref{fig:irsgroup}, based on which we estimate the combined channel of each group and consider a common reflection coefficient in the same group. Let $K$ denote the number of groups, with $1\leq K\leq M$. Without loss of generality, we assume equal size (number of IRS elements) of each group, which is given by $B=M/K$.\footnote{Without loss of generality, we assume that $B$ is an integer.} We further define the \emph{grouping ratio} as $\rho=1/B=K/M$, where $0<\rho\leq 1$; thus, a smaller grouping ratio implies more elements in each group. For illustration, we consider a rectangular-shaped IRS as shown in Fig. \ref{fig:irsgroup} with $\rho=1/4$, where there are $M_x$ and $M_y$ elements in each row and column, respectively. We further assume rectangular-shaped groups with $B_x$ and $B_y$ elements in each row and column, respectively, with $1\leq B_x\leq M_x$ and $1\leq B_y\leq M_y$. Therefore, the number of groups $K$ as well as the grouping ratio $\rho$ can be adjusted by varying $B_x$ and $B_y$. For ease of exposition, given any grouping, we assume the indices of IRS reflecting elements within each group are continuous, namely, the $\left((k-1)B+1\right)$-th to the $\left(kB\right)$-th elements belong to the $k$-th group, as illustrated in Fig. \ref{fig:irsgroup}. By considering a common reflection coefficient to the elements of each group, the IRS reflection coefficients can be re-expressed as 
\begin{align} 
 \boldsymbol{\phi}=\bar{\boldsymbol{\phi}}\otimes {\bf 1}_{B\times 1},
\end{align}
where $\bar{\boldsymbol{\phi}}=[\bar{\phi}_1,\cdots, \bar{\phi}_K]^T\in\mathbb{C}^{K\times 1}$ represents the IRS \emph{group} reflection coefficients, with $\bar{\phi}_k$ denoting the common reflection coefficient for the $k$-th group. With grouping, the concatenation of the BS-IRS channel, IRS reflection, and IRS-user channel in (\ref{eqn:hr1}) can be rewritten as
\begin{align}
\boldsymbol{h}_r =[{\boldsymbol{\nu}}_1,\dotsc,{\boldsymbol{\nu}}_M]\bar{\boldsymbol{\phi}}\otimes {\bf 1}_{B\times 1}=[\boldsymbol{\nu}'_1,\dotsc,\boldsymbol{\nu}'_K]\bar{\boldsymbol{\phi}}=\boldsymbol{V}'\bar{\boldsymbol{\phi}},
\end{align}
where $\boldsymbol{\nu}'_k=\sum_{b=1}^{B} {\boldsymbol{\nu}}_{b+(k-1)\times B}$ denotes the combined composite reflecting channel associated with the $k$-th IRS group, $k\in\mathcal{K}$, with $\mathcal{K}=\{1,\dotsc,K\}$, and $\boldsymbol{V}'\triangleq[\boldsymbol{\nu}'_1,\dotsc,\boldsymbol{\nu}'_K]\in\mathbb{C}^{N\times K}$ denotes the group composite channel matrix. It is worth noting that with our proposed grouping method, the size of the composite channel matrix associated with the IRS reflected link that needs to be estimated, $\boldsymbol{V}'$, is given by $N\times K$, which is generally smaller than that of the entire composite channel $\boldsymbol{V}$, $N\times M$, due to $K\leq M$. The maximum achievable rate based on the proposed IRS elements grouping is then given by  
\begin{align}
	r(\boldsymbol{p},\bar{\boldsymbol{\phi}})\!=\!\frac{1}{N\!+\!N_{\mathrm{CP}}}\!\sum_{n=0}^{N-1}\! \log_2\!\left(\!1\!+\!\frac{|\boldsymbol{f}_n^H\boldsymbol{h}_{d}\!+\!\boldsymbol{f}_n^H\boldsymbol{V}'\bar{\boldsymbol{\phi}}|^2p_n}{\Gamma\sigma^2} \!\right)\!,\label{eqn:rk2}
\end{align}
which is equivalent to (\ref{eqn:rk}) if $K=M$. 
Based on our proposed grouping method, we further propose a practical transmission protocol to approach the theoretical maximum achievable rate shown in (\ref{eqn:rk2}) in the next section.

\section{Transmission Protocol with IRS Elements Grouping}
\label{sec:ce}
\begin{figure}[!t]
 	\centering
 	\includegraphics[width=0.9\linewidth,keepaspectratio]{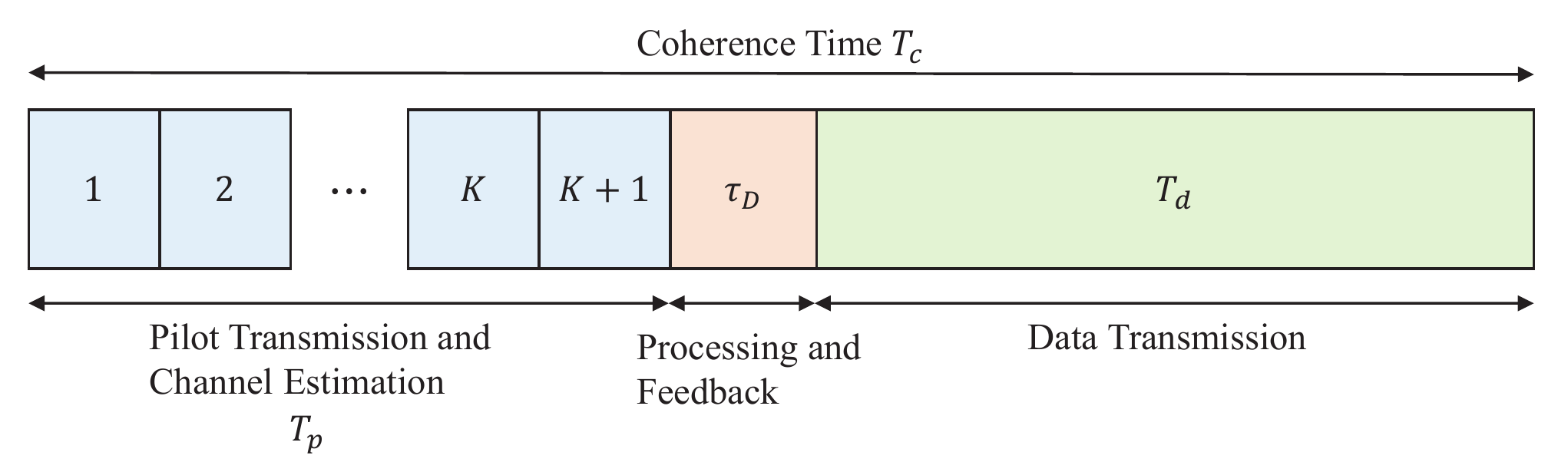}
 	\caption{Illustration of the proposed protocol with IRS elements grouping.}
 	\label{fig:Preamble}
 	\vspace{-6mm}
\end{figure}

In this section, we propose a practical transmission protocol based on the proposed grouping method, as illustrated in Fig. \ref{fig:Preamble}. Specifically, each channel coherence block consists of three transmission phases, with $T_c$ denoting the channel coherence time normalized to number of OFDM symbol durations. In the first phase, $K+1$ pilot symbols are transmitted from the user and reflected by the IRS, based on which the BS estimates the channel for both the BS-user direct link (i.e., $\boldsymbol{h}_d$) and the BS-IRS-user reflected link (i.e., $\boldsymbol{V}'$) for each IRS elements group. Then, based on the estimated channels, the BS computes the optimal transmit power allocation (i.e., $\boldsymbol{p}$) and IRS group reflection coefficients (i.e., $\bar{\boldsymbol{\phi}}$) that maximize the achievable rate, and feeds back $\boldsymbol{p}$ and $\bar{\boldsymbol{\phi}}$ to the user and the IRS controller, respectively. Finally, data transmission proceeds in the third phase based on the optimized $\boldsymbol{p}$ and $\bar{\boldsymbol{\phi}}$ and the estimated channels. 
In the following, we elaborate the three transmission phases in detail.

\subsection{Pilot Transmission and Channel Estimation}
Recall that the combined composite BS-IRS and IRS-user channel matrix $\boldsymbol{V}'\in\mathbb{C}^{N\times K}$ comprises all the reflecting channels associated with the respective IRS elements groups, which superpose with each other in the resultant BS-IRS-user link as $\boldsymbol{h}_r=\boldsymbol{V}'\bar{\boldsymbol{\phi}}$. To resolve their individual channels, we perform channel training based on the on/off state control of the IRS reflecting elements, which requires $T_p=K+1$ OFDM symbol durations as shown in Fig. \ref{fig:Preamble}. Specifically, let $\boldsymbol{x}_p\in\mathbb{C}^{N\times 1}$ denote the pilot signal sent at every symbol duration in the first phase, and $\boldsymbol{X}_p=\mathrm{diag}(\boldsymbol{x}_p)$. In the first symbol duration, all the IRS reflecting elements are switched off, i.e., $\bar{\boldsymbol{\phi}}={\bf 0}_{K\times 1}$. The equivalent baseband received signal in the frequency domain after CP removal is then given by 
\begin{align}\label{direct_pilot}
\boldsymbol{s}_0 = \boldsymbol{X}_p\boldsymbol{F}_N\boldsymbol{h}_d +\boldsymbol{n}_0,
\end{align} 
where $\boldsymbol{n}_0$ denotes the receiver noise with each element modelled by an independent CSCG random variable with mean zero and variance $\sigma^2$. Then, at the $(k+1)$-th pilot symbol, $k\in\mathcal{K}$, only the elements in the $k$-th group are switched on for full-reflection, while the remaining elements in other groups are switched off, i.e., $\bar{\boldsymbol{\phi}}=\boldsymbol{e}_k$, $k\in\mathcal{K}$. The equivalent baseband received signal in the frequency domain after CP removal is therefore given by 
\begin{align}\label{IRS_pilot}
\boldsymbol{s}_k = \boldsymbol{X}_p\boldsymbol{F}_N(\boldsymbol{h}_d +\boldsymbol{\nu}'_k) +\boldsymbol{n}_k,
\quad\quad k\in\mathcal{K},
\end{align}
where $\boldsymbol{n}_k$ denotes the receiver noise at the ($k+1$)-th pilot symbol duration,  with $\boldsymbol{n}_k\!\sim\!\mathcal{CN}\left(0,\sigma^2\boldsymbol{I}_N\right)$. Then, based on (\ref{direct_pilot}) and (\ref{IRS_pilot}), the BS-user channel and the composite reflecting channel can be obtained at the receiver by employing the least square (LS) estimation, which are expressed as \cite{lsest,lsest2}
\begin{align}
\hat{\boldsymbol{h}}_d &=\left[\left[\frac{1}{N}\boldsymbol{F}^H_N \boldsymbol{X}_p^{-1} \boldsymbol{s}_0\right]^T_{1:L},\; \boldsymbol{0}_{1\times (N-L)}\right]^T=\left[\left[\frac{1}{N}\boldsymbol{F}^H_N \boldsymbol{X}_p^{-1} \left(\boldsymbol{X}_p\boldsymbol{F}_N\boldsymbol{h}_d +\boldsymbol{n}_0\right) \right]^T_{1:L},\; \boldsymbol{0}_{1\times (N-L)}\right]^T\notag\\
&={\boldsymbol{h}}_d+\breve{\boldsymbol{n}}_0
,\label{direct_est}\\ 
\hat{\boldsymbol{\nu}}'_k &=\left[\left[\frac{1}{N}\boldsymbol{F}^H_N \boldsymbol{X}_p^{-1} \boldsymbol{s}_k-\hat{\boldsymbol{h}}_d\right]^T_{1:L_0},\; \boldsymbol{0}_{1\times (N-L_0)}\right]^T\notag\\
&=\left[\left[\frac{1}{N}\boldsymbol{F}^H_N \boldsymbol{X}_p^{-1} \left( \boldsymbol{X}_p\boldsymbol{F}_N(\boldsymbol{h}_d +\boldsymbol{\nu}'_k) +\boldsymbol{n}_k \right)-{\boldsymbol{h}}_d-\breve{\boldsymbol{n}}_0\right]^T_{1:L_0},\; \boldsymbol{0}_{1\times (N-L_0)}\right]^T\notag\\
&={\boldsymbol{\nu}}'_k +\breve{\boldsymbol{n}}_k-\breve{\boldsymbol{n}}_0
, \quad k\in\mathcal{K},\label{reflect_est}
\end{align}
where $\breve{\boldsymbol{n}}_0=\left[\left[\frac{1}{N}\boldsymbol{F}^H_N \boldsymbol{X}_p^{-1} \boldsymbol{n}_0 \right]^T_{1:L},\; \boldsymbol{0}_{1\times (N-L)}\right]^T$ and 
$\breve{\boldsymbol{n}}_k=\left[\left[\frac{1}{N}\boldsymbol{F}^H_N \boldsymbol{X}_p^{-1} \boldsymbol{n}_k \right]^T_{1:L_0},\; \boldsymbol{0}_{1\times (N-L_0)}\right]^T$ for $k\in\mathcal{K}$.\footnote{We may need to assume that $L\le L_0$ for (\ref{reflect_est}) to avoid improper truncation of ${\boldsymbol{h}}_d$. This is highly likely in practice, since the signal path of the BS-IRS-user reflected channel is typically longer than the BS-user direct channel.} Moreover, based on (\ref{direct_est}) and (\ref{reflect_est}), the average mean square error (MSE) of channel estimation on each SC can be derived as
\begin{align}
\varepsilon &=\frac{1}{N}  \cdot {\mathbb E}\left\{  \left\|
\boldsymbol{h}_d +{\boldsymbol{V}}'{\bar{\boldsymbol{\phi}}}-\hat{\boldsymbol{h}}_d -\hat{\boldsymbol{V}}' {\bar{\boldsymbol{\phi}}}
\right\|^{2}
\right\}\nonumber\\
&=\frac{1}{N}  \cdot {\mathbb E}\left\{  \left\|
\breve{\boldsymbol{n}}_0
+\left[\breve{\boldsymbol{n}}_1-\breve{\boldsymbol{n}}_0,\ldots, \breve{\boldsymbol{n}}_K-\breve{\boldsymbol{n}}_0 \right] {\bar{\boldsymbol{\phi}}}
\right\|^{2}
\right\}\nonumber\\
&\le \frac{\sigma^2(K+1)L+\sigma^2KL_0 }{P_{t}},\label{MSE}
\end{align}
where $P_t$ denotes the total transmit power during channel training. It can be observed from \eqref{MSE} that the MSE is inversely proportional to the training SNR and proportional to the number of groups $K$, i.e., we can reduce the channel estimation error by increasing the transmit power during the training phase and/or the group size (i.e., smaller $K$).
Meanwhile, it is worth noting that, with the proposed IRS elements grouping, the pilot training period $T_p$ is reduced to $K+1$ OFDM symbols as compared to $M+1$ without the grouping, if $K<M$.\footnote{It is worth pointing out that during the preparation of this work, we became aware of a related work \cite{Mishra2019Channel} on the channel estimation for IRS-enhanced systems, which independently proposed a sequential channel estimation method under the frequency-flat channel that requires $M+1$ pilot symbols in the training phase. This method can be considered as a special case of our proposed scheme without the grouping.}

\subsection{Processing and Feedback}
In the second phase, based on the estimated CSI $\hat{\boldsymbol{h}}_d$ and $\hat{\boldsymbol{V}}'$ at the BS, the transmit power allocation and IRS group reflection coefficients are optimized at the BS to maximize the achievable rate, which is given by
\begin{align}
r(\boldsymbol{p},\bar{\boldsymbol{\phi}}|\hat{\boldsymbol{h}}_d, \hat{\boldsymbol{V}}')=\left(1-\frac{T_{\mathrm{p}}+\tau_{D}}{T_{c}}\right)\frac{1}{N+N_{\mathrm{CP}}} \sum_{n=0}^{N-1}\log_2 \left(1+\frac{\left|\boldsymbol{f}_n^H \hat{\boldsymbol{h}}_d+\boldsymbol{f}_n^H \hat{\boldsymbol{V}}'{\bar{\boldsymbol{\phi}}}
	\right|^2 p_n }{\Gamma\sigma^2}\right),\label{eqn:r2}
\end{align}
where $\tau_D$ denotes the processing and feedback delay. For simplicity, we ignore the estimation error during parameter optimization, while the effect of estimation error will be taken into account in the simulations. Let $\boldsymbol{p}^{\star}$ and $\bar{\boldsymbol{\phi}}^{\star}$ denote the set of optimal transmit power allocation and IRS reflection coefficients, respectively. The details on how to obtain $\boldsymbol{p}^{\star}$ and $\bar{\boldsymbol{\phi}}^{\star}$ will be presented in Section \ref{sec:pfsol}. Afterwards, the BS feeds back $\boldsymbol{p}^{\star}$ to the user and $\bar{\boldsymbol{\phi}}^{\star}$ to the IRS controller, respectively.

\subsection{Data Transmission}
Finally, with the optimized transmit power allocation $\boldsymbol{p}^{\star}$ available at both the BS and the user as well as the optimized IRS group reflection coefficients $\bar{\boldsymbol{\phi}}^{\star}$ available at the BS and the IRS, all IRS elements are switched on and data transmission proceeds with the optimized design parameters. 
Note that due to the presence of channel estimation error, which can be viewed equivalently as additional interference at the receiver, the achievable rate for actual data transmission will be lower than that obtained in \eqref{eqn:r2}.
Furthermore, it is worth noting that there exists a fundamental \emph{trade-off} between the training overhead versus the IRS reflection coefficient design flexibility. Specifically, as the number of groups $K$ and consequently the grouping ratio $\rho$ decreases, the estimation error decreases while the effective data transmission time increases, while the reduced number of elements in $\bar{\boldsymbol{\phi}}$ leads to less flexibility in the IRS reflection coefficient design and thereby potentially smaller achievable rate in the data transmission phase. Nevertheless, as the channel correlation among the IRS elements in the same group becomes stronger, it is anticipated that such design flexibility loss will be compensated by the saved data transmission time. Moreover, the optimal grouping ratio is also dependent on the channel coherence time. A comprehensive illustration of this trade-off will be given later in Section \ref{sec:sim} via numerical results.

\section{Achievable Rate Maximization via Joint Power Allocation and Reflection Coefficient Optimization}
\label{sec:pfsol}
\subsection{Problem Formulation}
Based on the transmission protocol proposed in Section \ref{sec:ce} as well as given IRS elements grouping $K$ and channel estimates $\hat{\boldsymbol{h}}_d$ and $\hat{\boldsymbol{V}}'$, we aim to maximize the achievable rate shown in (\ref{eqn:r2}) by jointly optimizing the transmit power allocation and the IRS reflection coefficients. Therefore, we formulate the following optimization problem (by omitting the constant terms in (\ref{eqn:r2}) for brevity)
\begin{align}
\mathrm{(P1)}:\mathop{\mathtt{maximize~~}}_{\boldsymbol{p},\bar{\boldsymbol{\phi}}~}  &~~\sum_{n=0}^{N-1} \log_2\left(1\!+\!\frac{\left|\boldsymbol{f}_n^H\hat{\boldsymbol{h}}_{d}\!+\!\boldsymbol{f}_n^H\hat{\boldsymbol{V}}'\bar{\boldsymbol{\phi}}\right|^2p_n}{\Gamma\sigma^2} \right)	\nonumber\\
	\mathtt{subject\; to}
&~~\sum_{n=0}^{N-1} p_{n}\leq P, \label{eqn:const1}\\
&~~p_n \geq0, \quad \forall n \in \mathcal{N}, \label{eqn:const2}\\
&~~|\bar{\phi}_k|\leq 1, \quad \forall k\in \mathcal{K}.\label{eqn:const3}
\end{align}
Problem (P1) is a non-convex optimization problem. Particularly, it can be shown that the objective function of (P1) is non-concave over $\bar{\boldsymbol{\phi}}$; moreover, the variables $\bar{\boldsymbol{\phi}}$ and $\boldsymbol{p}$ are coupled in the objective function, which makes their joint optimization difficult. To overcome the above challenges, in the following, we propose an alternating optimization algorithm to find an approximate solution to (P1), by iteratively optimizing one of $\boldsymbol{p}$ and $\bar{\boldsymbol{\phi}}$ with the other fixed at each time. In addition, we devise a customized method to obtain an initial solution of $\bar{\boldsymbol{\phi}}$, denoted by $\bar{\boldsymbol{\phi}}_0$, as the starting point of the proposed alternating optimization algorithm. 

\subsection{Power Allocation Optimization Given IRS Coefficients}
\label{sec:pa}
Note that given a set of IRS coefficients $\bar{\boldsymbol{\phi}}$ with the CSI estimates, the CFR can be estimated as $\hat{\boldsymbol{v}}=\boldsymbol{F}_N\left(\hat{\boldsymbol{h}}_d+\hat{\boldsymbol{V}}'\bar{\boldsymbol{\phi}}\right)$. The optimal BS transmit power allocation $\boldsymbol{p}$ is thus given by the well-known water-filling (WF) solution \cite{goldsmith}, i.e.,
\begin{equation}\label{eqn:p1}
	p_n=\left(\frac{1}{c_u}-\frac{1}{c_n}\right)^+,\quad\forall n\in\mathcal{N},
\end{equation}
where $\left(x\right)^+\triangleq\max\left(0,x\right)$, $c_n=|\hat{v}_n|^2/(\Gamma\sigma^2)$ is the effective channel-to-noise power ratio (CNR) for SC $n$, and $c_u$ is the cut-off CNR that satisfies
\begin{align}
	\sum_{n=0}^{N-1} \left(\frac{1}{c_u}-\frac{1}{c_n}\right)^+=P.\label{eqn:p2}
\end{align} 

\subsection{IRS Coefficient Optimization Given Power Allocation}
Given power allocation, Problem (P1) is simplified as 
\begin{align}
\mathrm{(P1.1)}:\mathop{\mathtt{maximize~~}}_{\bar{\boldsymbol{\phi}}~}  &~\sum_{n=0}^{N-1} \log_2\left(\!1\!+\!\frac{|\boldsymbol{f}_n^H\hat{\boldsymbol{h}}_{d}\!+\!\boldsymbol{f}_n^H\hat{\boldsymbol{V}}'\bar{\boldsymbol{\phi}}|^2p_n}{\Gamma\sigma^2} \!\right)	\nonumber\\
	\mathtt{subject\; to}
&~~|\bar{\phi}_k|\leq 1, \quad \forall k\in \mathcal{K}.
\end{align}
It can be shown that (P1.1) is a non-convex optimization problem. In the following, we adopt the successive convex approximation (SCA) technique \cite{sca} to obtain a locally optimal solution to (P1.1). First, by introducing a set of auxiliary variables $y_n$'s, $a_n$'s, and $b_n$'s, we transform (P1.1) into the following equivalent problem
\begin{align}
\mathrm{(P1.1')}:\mathop{\mathtt{maximize~~}}_{\bar{\boldsymbol{\phi}},\{\!y_n\!\},\{\!a_n\!\},\{\!b_n\!\}}  &~\sum_{n=0}^{N-1} \log_2\left(1+\frac{y_np_n}{\Gamma\sigma^2} \right)	\nonumber\\
	\mathtt{subject\; to}
&~|\bar{\phi}_k|\leq 1, \quad \forall k\in \mathcal{K},\label{eqn:phim}\\
&~a_n\!=\!\Re\{\boldsymbol{f}_n^H\hat{\boldsymbol{h}}_{d}\!+\!\boldsymbol{f}_n^H\hat{\boldsymbol{V}}'\bar{\boldsymbol{\phi}}\},\;\forall n\!\in\!\mathcal{N},\label{eqn:an}\\
&~b_n\!=\!\Im\{\boldsymbol{f}_n^H\hat{\boldsymbol{h}}_{d}\!+\!\boldsymbol{f}_n^H\hat{\boldsymbol{V}}'\bar{\boldsymbol{\phi}}\},\;\forall n\!\in\!\mathcal{N},\label{eqn:bn}\\
&~y_n\leq a_n^2+b_n^2, \quad\forall n\in\mathcal{N}.\label{eqn:const4}
\end{align}
Define $\tilde{f}_n(a_n,b_n)\triangleq a_n^2+b_n^2$, which is a convex and differentiable function over $a_n$ and $b_n$. Thus, given any $\tilde{a}_n$ and $\tilde{b}_n$, the first-order approximation of $\tilde{f}_n(a_n,b_n)$ at the point $(\tilde{a}_n,\tilde{b}_n)$ serves as a lower bound to it, i.e.,
\begin{align}
	\tilde{f}_n(a_n,b_n)\geq\tilde{a}_n^2+\tilde{b}_n^2+2\tilde{a}_n(a_n-\tilde{a}_n)+2\tilde{b}_n(b_n-\tilde{b}_n)\triangleq f_n(a_n,b_n),
\end{align}
where equality holds if and only if $\tilde{a}_n=a_n$ and $\tilde{b}_n=b_n$. Note that $f_n(a_n,b_n)$ is an affine function over $a_n$ and $b_n$, which also has the same gradient over $a_n$ and $b_n$ as $\tilde{f}_n(a_n,b_n)$ at the point $(\tilde{a}_n,\tilde{b}_n)$.

Next, we consider the following optimization problem
\begin{align}
\mathrm{(P1.2)}:\mathop{\mathtt{maximize~~}}_{\bar{\boldsymbol{\phi}},\{\!y_n\!\}\!,\{\!a_n\!\}\!,\{\!b_n\!\}\!~}  &~\sum_{n=0}^{N-1} \log_2\left(1+\frac{y_np_n}{\Gamma\sigma^2} \right)	\nonumber\\
	\mathtt{subject\; to}
&~(\ref{eqn:phim}),(\ref{eqn:an}),(\ref{eqn:bn})\nonumber\\
&~y_n\leq f_n(a_n,b_n),\quad\forall n\in\mathcal{N}.\label{eqn:const7}
\end{align}
Problem (P1.2) is a convex optimization problem, which can be solved efficiently via existing software, e.g., CVX \cite{cvx}. The computation complexity for solving (P1.2) can be obtained by considering the complexity analysis of the interior-point method. Specifically, (P1.2) involves $K$ second-order cone (SOC) constraints of size 2, $4N$ linear inequality constraints ($2N$ linear equality constraints) of size $2K+1$, and $N$ linear inequality constraints of size $3$, with a total number of $2K+3N$ optimization variables. The total complexity for solving (P1.2) is thus given by $\sqrt{4N(2K+1)+3N+2K}(2K+3N)(4N(2K+1)^3+27N+(2K+3N)(4N(2K+1)^2+9N)+4K+(2K+3N)^2)$ \cite{complexity}, i.e., $\mathcal{O}(K^{4.5}N^{3.5})$. Therefore, an approximate solution to (P1.1$'$) and thus (P1.1) can be obtained by successively updating $\{\tilde{a}_n\}$ and $\{\tilde{b}_n\}$ based on the optimal solution to (P1.2), which is summarized in Algorithm 1. It can be shown that monotonic convergence of Algorithm 1 is guaranteed, and the obtained solution is a locally optimal solution to (P1.1) \cite{sca}. 

\begin{algorithm}[!t]
\DontPrintSemicolon
\caption{IRS Coefficient Optimization Given Power Allocation via SCA}
\KwIn{$\hat{\boldsymbol{h}}_d$, $\hat{\boldsymbol{V}}'$, $\boldsymbol{p}$, $\Gamma$, $\sigma^2$, $N$, $K$, $\tilde{\boldsymbol{\phi}}$.}
\KwOut{$\bar{\boldsymbol{\phi}}$.} 
Set $\tilde{a}_n=\Re\{\boldsymbol{f}_n^H\hat{\boldsymbol{h}}_{d}\!+\!\boldsymbol{f}_n^H\hat{\boldsymbol{V}}'\tilde{\boldsymbol{\phi}}\}$, $\tilde{b}_n=\Im\{\boldsymbol{f}_n^H\hat{\boldsymbol{h}}_{d}\!+\!\boldsymbol{f}_n^H\hat{\boldsymbol{V}}'\tilde{\boldsymbol{\phi}}\}$, $\forall n\in\mathcal{N}$.

\Repeat{\upshape the objective value of (P1.1) with the obtained $\bar{\boldsymbol{\phi}}$ reaches convergence.}
   	{Find the optimal solution of $\{a_n\}$, $\{b_n\}$, and $\bar{\boldsymbol{\phi}}$ to (P1.2) via CVX with given $\{\tilde{a}_n\}$, $\{\tilde{b}_n\}$, and $\boldsymbol{p}$.
   	
   	$\tilde{a}_n=a_n$, $\tilde{b}_n=b_n$, $\forall n\in\mathcal{N}$.}
\end{algorithm}

To summarize, the overall iterative algorithm to solve (P1) is given in Algorithm~2. It is worth noting that starting from an initial point denoted by $\bar{\boldsymbol{\phi}}_0$, the initial value $\tilde{\boldsymbol{\phi}}$ for Algorithm~1 in each iteration of Algorithm~2 is set as the obtained $\bar{\boldsymbol{\phi}}$ in the previous iteration. It can be shown that the objective value of (P1) is non-decreasing over each iteration of Algorithm~2, which is also upper-bounded by a finite value. Therefore, Algorithm~2 is guaranteed to converge. Moreover, the obtained solution to (P1) can be shown to be at least a locally optimal solution based on \cite{conv}. Note that the performance of Algorithm~2 is critically dependent on the choice of the initial IRS reflection coefficients $\bar{\boldsymbol{\phi}}_0$. In the following subsection, we propose a customized method for finding $\bar{\boldsymbol{\phi}}_0$ efficiently. 

\begin{algorithm}[!t]
\DontPrintSemicolon
\caption{Alternating Optimization for Solving (P1)}
\KwIn{$\hat{\boldsymbol{h}}_d$, $\hat{\boldsymbol{V}}'$, $P$, $\Gamma$, $\sigma^2$, $N$, $K$, $\bar{\boldsymbol{\phi}}=\bar{\boldsymbol{\phi}}_0$.}
\KwOut{$\boldsymbol{p}^{\star}$, $\bar{\boldsymbol{\phi}}^{\star}$.}
\Repeat{\upshape the objective value of (P1) with the obtained $\boldsymbol{p}$ and $\bar{\boldsymbol{\phi}}$ reaches convergence.}
   {Fixing the IRS coefficients $\bar{\boldsymbol{\phi}}$, find the WF power allocation $\boldsymbol{p}$ according to (\ref{eqn:vn}), (\ref{eqn:p1}), and (\ref{eqn:p2}).
   
   Fixing the power allocation $\boldsymbol{p}$, given initial $\tilde{\boldsymbol{\phi}}=\bar{\boldsymbol{\phi}}$, update the IRS coefficients $\bar{\boldsymbol{\phi}}$ via Algorithm 1.}
   
$\boldsymbol{p}^{\star}=\boldsymbol{p}$, $\bar{\boldsymbol{\phi}}^{\star}=\bar{\boldsymbol{\phi}}$.
\end{algorithm}

\subsection{Initialization Method}
\label{sec:init}
Note that the IRS is able to increase the link rate mainly due to the increased effective channel power between the BS and the user, by creating an additional strong CIR via the BS-IRS-user channel that can constructively combine with that of the BS-user direct channel. Motivated by this, we propose to design the initial value of $\bar{\boldsymbol{\phi}}$, i.e., $\bar{\boldsymbol{\phi}}_0$, by maximizing the effective channel power from the BS to the user, which is given by $\left\|\hat{\boldsymbol{h}}_d+\hat{\boldsymbol{V}}'\bar{\boldsymbol{\phi}}\right\|^2$. Therefore, we formulate the following optimization problem
\begin{align}
\mathrm{(P2)}:~\mathop{\mathtt{maximize~~}}_{\bar{\boldsymbol{\phi}}~}  &~\left\|\hat{\boldsymbol{h}}_d+\hat{\boldsymbol{V}}'\bar{\boldsymbol{\phi}}\right\|^2	\nonumber\\
	\mathtt{subject\; to}
&~~|\bar{\phi}_k|^2\leq 1, \quad \forall k\in\mathcal{K}.\label{eqn:const5}
\end{align}
Note that Problem (P2) is a non-convex quadratically constrained quadratic problem (QCQP), for which we apply the semidefinite relaxation (SDR) \cite{sdr} technique to obtain an approximate solution for it, as follows. Define $\boldsymbol{A}\triangleq\left(\hat{\boldsymbol{V}}'\right)^H\hat{\boldsymbol{V}}'$ and $\boldsymbol{u}\triangleq\left(\hat{\boldsymbol{V}}'\right)^H\hat{\boldsymbol{h}}_d$, Problem (P2) is thus equivalent to 
\begin{align}
	~\mathop{\mathtt{maximize~~}}_{\bar{\boldsymbol{\phi}}~} &~\bar{\boldsymbol{\phi}}^H\boldsymbol{A}\bar{\boldsymbol{\phi}}+\bar{\boldsymbol{\phi}}^H\boldsymbol{u}+\boldsymbol{u}^H\bar{\boldsymbol{\phi}} \\
	\mathtt{subject\; to}&~~|\bar{\phi}_k|^2\leq 1, \quad \forall k\in\mathcal{K}.
\end{align}
Note that $\bar{\boldsymbol{\phi}}^H\boldsymbol{A}\bar{\boldsymbol{\phi}}=\mathrm{Tr}\left(\bar{\boldsymbol{\phi}}^H\boldsymbol{A}\bar{\boldsymbol{\phi}}\right)=\mathrm{Tr}\left(\bar{\boldsymbol{\phi}}\bar{\boldsymbol{\phi}}^H\boldsymbol{A}\right)$; similarly, $\bar{\boldsymbol{\phi}}^H\boldsymbol{u}=\mathrm{Tr}\left(\boldsymbol{u}\bar{\boldsymbol{\phi}}^H\right)$ and $\boldsymbol{u}^H\bar{\boldsymbol{\phi}}=\mathrm{Tr}\left(\bar{\boldsymbol{\phi}}\boldsymbol{u}^H\right)$ hold. By defining $\boldsymbol{w}=[\bar{\boldsymbol{\phi}},\boldsymbol{u}]^T$ and $\boldsymbol{W}=\boldsymbol{w}\boldsymbol{w}^H$, we transform (P2) into the following problem
\begin{align}
\mathrm{(P2-SDR)}:~\mathop{\mathtt{maximize~~}}_{\boldsymbol{W}~}  ~&~\mathrm{Tr}\left(\boldsymbol{W}\boldsymbol{M}\right)\nonumber\\
	\mathtt{subject\; to}~
&~\boldsymbol{W}_{k,k}\leq1, \quad \forall k\in\mathcal{K},\\
&~\boldsymbol{W}_{k,k}=|u_{k-K}|, \quad \forall k\!-\!K\!\in\!\mathcal{K},\\
&~\boldsymbol{W}\succeq\boldsymbol{0},\label{eqn:const7}
\end{align}
where $\boldsymbol{M}=[\boldsymbol{A},\boldsymbol{I}_K; \boldsymbol{I}_K,\boldsymbol{0}_{K\times K}]$, and the constraint in (\ref{eqn:const7}) ensures $\boldsymbol{W}$ is positive semidefinite. Note that (P2) can be shown to be equivalent to (P2-SDR) with the additional constraint of $\mathrm{rank}(\boldsymbol{W})=1$. Problem (P2-SDR) is a convex semidefinite program (SDP), which can be solved efficiently via existing software, e.g., CVX \cite{cvx}, with a complexity of $\mathcal{O}(K^{4.5})$ \cite{sdr}. Let $\boldsymbol{W}^{\star}$ denote the optimal solution to (P2-SDR). If $\mathrm{rank}\left(\boldsymbol{W}^{\star}\right)\!=\!1$, the relaxation from (P2) to (P2-SDR) is tight and the optimal $\bar{\boldsymbol{\phi}}$ to Problem (P2) can be obtained as $\bar{\boldsymbol{\phi}}^{\star}\!=\!\boldsymbol{U}\mathrm{Diag}\left(\boldsymbol{\Lambda}^{\frac{1}{2}}\right)$, where $\boldsymbol{U}\boldsymbol{\Lambda}\boldsymbol{U}^H$ is the eigenvalue decomposition (EVD) of the upper left $K\!\times\! K$ submatrix of $\boldsymbol{W}^{\star}$, denoted by $\boldsymbol{W}_s^{\star}$. On the other hand, if $\mathrm{rank}\left(\boldsymbol{W}^{\star}\right)\!>\!1$, the optimal objective value of Problem (P2-SDR) serves as an upper bound to that of Problem (P2) and additional processing is required to construct a rank-one solution according to $\boldsymbol{W}^{\star}$. In particular, we consider a customized Gaussian randomization method \cite{sdrrand} to find an approximate solution to Problem (P2). To enhance the performance of the proposed algorithm, a number (denoted by $Q$) of $\hat{\boldsymbol{\phi}}$'s are generated based on $\boldsymbol{W}^{\star}$, from which the one that yields the largest objective value of Problem (P2) is selected. The overall algorithm for solving (P2) is summarized in Algorithm 3, where the output $\bar{\boldsymbol{\phi}}$ of Algorithm 3 is then set as the initial $\bar{\boldsymbol{\phi}}_0$ for Algorithm 2.

\begin{algorithm}[t]
\caption{Algorithm for Solving (P2)}

\KwIn{$\hat{\boldsymbol{h}}_d$, $\hat{\boldsymbol{V}}'$, $K$, $Q$.}
\KwOut{$\bar{\boldsymbol{\phi}}$.}
Solve (P2-SDR) via CVX and obtain the optimal solution $\boldsymbol{W}^{\star}$. 

Obtain the submatrix $\boldsymbol{W}_s^{\star}$ by $[\boldsymbol{W}_s^{\star}]_{i,j}=[\boldsymbol{W}^{\star}]_{i,j}$, $i\in\mathcal{K}$, $j\in\mathcal{K}$. 

Compute the EVD of $\boldsymbol{W}_s^{\star}$ by $\boldsymbol{W}_s^{\star}=\boldsymbol{U}\boldsymbol{\Lambda}\boldsymbol{U}^H$.

\eIf{$\mathrm{rank}\left(\boldsymbol{W}^{\star}\right)=1$}
	{$\bar{\boldsymbol{\phi}}=\boldsymbol{U}\mathrm{Diag}\left(\boldsymbol{\Lambda}^{\frac{1}{2}}\right)$. }
{
\For{$q=1$ \KwTo $Q$}{
   Generate $\tilde{\boldsymbol{r}}^{(q)}\sim\mathcal{CN}\left(\boldsymbol{0},\boldsymbol{I}_{K}\right)$. Obtain $\hat{\boldsymbol{\phi}}^{(q)}=e^{j\arg\left(\boldsymbol{U}\boldsymbol{\Lambda}^{\frac{1}{2}}\tilde{\boldsymbol{r}}^{(q)}\right)}$. 
   
    Compute the corresponding channel power $P_h^{(q)}=\left\|\hat{\boldsymbol{h}}_d+\hat{\boldsymbol{V}}'\hat{\boldsymbol{\phi}}^{(q)}\right\|^2	$.
}

$q^{\star}=\argmax{q=1,\dotsc,Q} P_h^{(q)}$, $\bar{\boldsymbol{\phi}}=\hat{\boldsymbol{\phi}}^{(q^{\star})}$.
}
\end{algorithm}

Note that the complexity for solving the SDP in (P2-SDR) could be practically high for large $K$. To reduce the complexity of the initialization method, we further propose a low-complexity suboptimal method to solve (P2), which is based on successive alignment (SA) of the reflection coefficients. Specifically, starting from a set of initial reflection coefficients with random phase and unit amplitude, the phase of one reflecting element is optimized at a time with all the other reflection coefficients being fixed, and the reflection coefficients of all reflecting elements are updated sequentially in one iteration. The optimal reflection coefficient at the $i$-th reflecting element with $\{\bar{\phi}_k\}_{k\in\mathcal{K},k\neq i}$ being fixed can be obtained as
\begin{align}
	\bar{\phi}_i^{\star}=e^{-j\arg\left\{\left(\boldsymbol{h}_d^H+\sum_{k\neq i,k\in\mathcal{K}}\boldsymbol{\nu}'^H_k\bar{\phi}_k^*\right)\boldsymbol{\nu}'_i\right\}},\quad i\in\mathcal{K}.
\end{align} 
The complexity for updating all reflection coefficients in one iteration is given by $\mathcal{O}(KN)$. Let $I_{SA}$ denote the number of iterations that the SA is performed before proceeding to the alternating optimization in Algorithm 2, the total complexity for this initialization method is thus $\mathcal{O}(I_{SA}KN)$. It is observed that the SA initialization is able to obtain similar performance with the SDR initialization, yet with generally much lower complexity, as will be shown in Section \ref{sec:sim}.

\section{Numerical Results}
\label{sec:sim}
In this section, we evaluate the performance of our proposed protocol and algorithm designs via numerical results, for both downlink and uplink communications. We consider an OFDM system with $N=64$ SCs. For the BS-user direct link, we consider a Rayleigh fading channel with a maximum delay spread of $L=16$ taps, where each tap coefficient is modeled as a zero-mean CSCG random variable with a uniform power delay profile. For the BS-IRS-user composite link, note that the deployment location of the IRS can be chosen to favor line-of-sight (LoS) transmission between the IRS and the BS/user. Hence, we model the BS-IRS channel with a maximum delay spread of $L_1=4$ taps and the IRS-user channel with $L_2=13$ taps (e.g., a rich-scattering indoor environment), with the first tap of each channel being the LoS signal path and the remaining taps characterizing the non-LoS (NLoS) paths. The CP length is thus set as $N_{\mathrm{CP}}=16$. The NLoS channel taps are modeled by Rayleigh fading similar to the BS-user direct link. Let $\zeta_{BI}$ and $\zeta_{Iu}$ denote the power ratio of the LoS component to the NLoS component for the BS-IRS and IRS-user channels, respectively. Hence, $\zeta_{BI}\rightarrow\infty$ indicates an LoS channel while $\zeta_{BI}\rightarrow 0$ indicates a Rayleigh fading channel, similarly for $\zeta_{Iu}$. For illustration, we set $\zeta_{BI}=3$ dB and $\zeta_{Iu}=-20$ dB. For the LoS path, as the IRS array size is practically much smaller than the distance between the IRS and the BS/user (typical metasurface cell dimension is around $5$ mm \cite{metasur,metasurcell} while typical link distance is in the range of tens to hundreds of meters), the channel gains for all elements of the IRS are approximately the same while their phases are correlated depending on the physical layout. For ease of exposition, we assume the IRS is placed along the $x-y$ plane and perpendicular to the ground ($x-z$ plane).
The rays arriving at the IRS are assumed to be parallel for all elements, with a common angle of arrival (AoA) composed of an elevation angle of $\psi_e$
and an azimuth angle of $\psi_a$.
In the following results with varying $M$, we fix $M_x=5$ and increase $M_y$ linearly with $M$, unless otherwise stated. Let $(m_x,m_y)$ denote the location of an IRS element, with $1\leq m_x\leq M_x$ and $1\leq m_y \leq M_y$. Let $\omega(m_x,m_y)$ denote the phase offset of the IRS-user link at $(m_x,m_y)$ with respect to that at $(1,1)$, thus we have $\omega(m_x,m_y)=\frac{2\pi}{\lambda}((m_x-1)d\sin\psi_e\sin\psi_a+(m_y-1)d\cos\psi_e))$, where $d=0.01$ m denotes the IRS elements separation and $\lambda=0.0857$ m denotes the carrier wavelength (corresponding to a 3.5 GHz carrier frequency). The phase offset for the LoS path of the BS-IRS link can be obtained similarly. As a result, the phase difference of the LoS paths for all IRS elements in the BS-IRS link and IRS-user link are fixed. The average channel power of the direct link is given by $P_d=\mathbb{E}[\|\boldsymbol{h}_d\|^2]$, and that of the reflected link at each reflecting element is given by $P_{r,m}=\mathbb{E}[\|\boldsymbol{\nu}_m\|^2]$, $\forall m\in\mathcal{M}$. For convenience, we normalize the direct link power as $P_d=1$, and the direct link (reference) SNR is thus given by $\gamma_d=P/(N\sigma^2)$. Let $\alpha=P_{r,m}/P_d$ denote the average power ratio of the reflected channel at one reflecting element to the direct link. Hence, $\alpha\rightarrow 0$ indicates that the user is located sufficiently far from the IRS, thus its channel with the BS is dominated by the BS-user direct link; while on the other hand, $\alpha\gg 0$ indicates that the user is located in close vicinity of the IRS. The SNR gap is set as $\Gamma=8.8$ dB \cite{snrgap}, and the number of randomizations in Algorithm 3 is chosen as $Q=50$. All the results are averaged over 100 independent channel realizations.

\subsection{Performance of Proposed Algorithms with Perfect CSI}
In this subsection, we evaluate the performance of the proposed algorithms in Section \ref{sec:pfsol}. For the purpose of exposition, we assume perfect CSI is available without any grouping (i.e., $K=M$), and ignore any overhead pertaining to channel estimation in this part. The achievable rate in this subsection is thus computed using (\ref{eqn:rk}).

For comparison, we consider the following benchmark schemes:
\begin{enumerate}
	\item \textbf{Iterative (amplitude=1)}: In this scheme, we force the amplitudes of the obtained reflection coefficients to unity at the end of Algorithm 2, and then obtain the corresponding WF transmit power allocation and achievable rate.
	\item \textbf{Channel Power Maximization (CPM)}: In this scheme, we adopt the IRS coefficients as $\bar{\boldsymbol{\phi}}_0$ obtained via the initialization methods based on the CPM proposed in Section \ref{sec:init}, and the WF transmit power allocation based on $\bar{\boldsymbol{\phi}}_0$.
	\item \textbf{Random Phase}: We assume each IRS coefficient has a random phase independently and uniformly distributed in $[0,2\pi]$ and the maximum amplitude, based on which we obtain the WF transmit power allocation. 
	\item \textbf{Without IRS}: We consider the WF transmit power allocation and achievable rate based on the BS-user direct link only.
\end{enumerate}

\begin{figure}[t]
    \centering
    \includegraphics[width=0.6\linewidth, keepaspectratio]{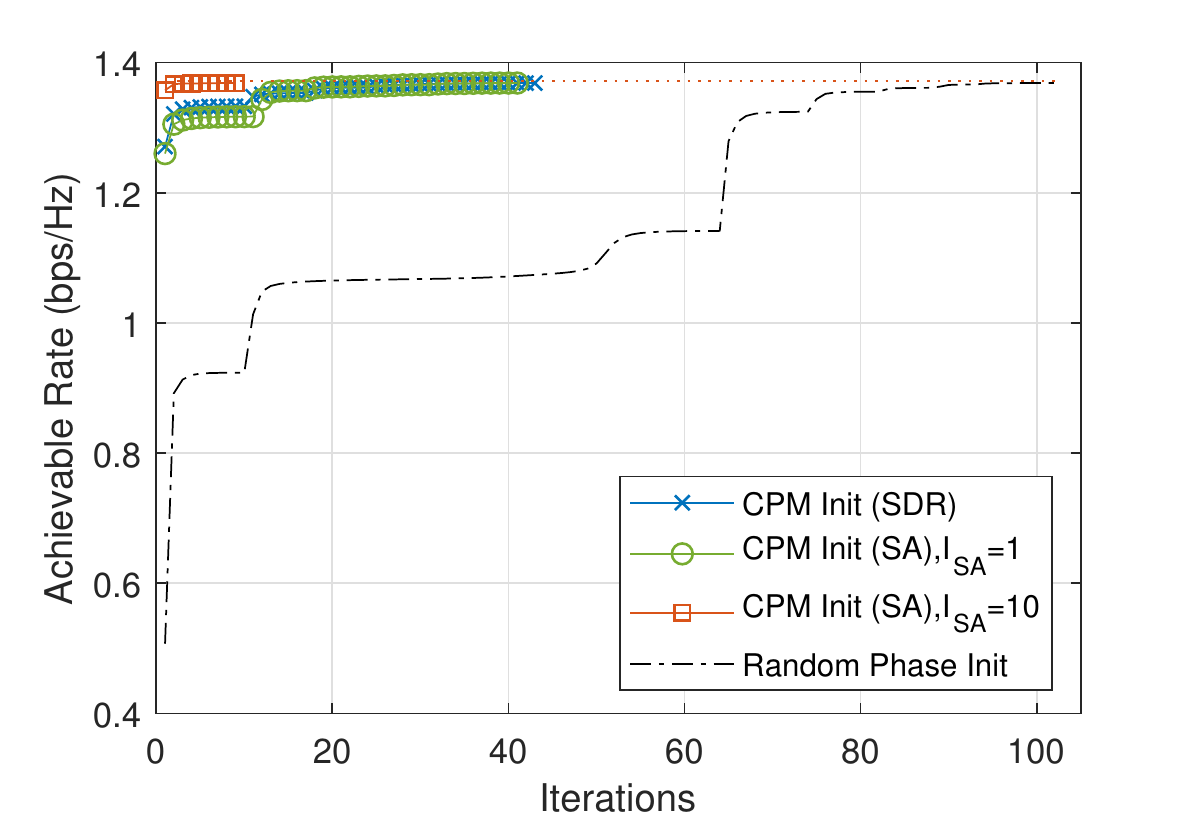}
    \vspace{-3mm}
    \caption{Convergence behavior of Algorithm 2.}
    \label{fig:ini_comp}
    \vspace{-7mm}
\end{figure}
First, we evaluate the convergence behavior of Algorithm 2. The number of reflecting elements is set as $M=20$, and the power ratio of the reflected link at one reflecting element to the direct link is set as $\alpha=0.05$. For comparison with the CPM-based initialization methods proposed in Section \ref{sec:init}, we consider a benchmark initialization method with random phases. Fig. \ref{fig:ini_comp} shows the achievable rate over iterations at $\gamma_d=5$ dB for a random channel realization. Monotonic convergence is observed for all initialization methods, which is consistent with our discussions in Section \ref{sec:pfsol}. Moreover, it is observed that the SA-based CPM initialization method with $I_{SA}=1$ has similar performance and convergence rate compared to that of the SDR-based CPM initialization counterpart (i.e., $41$ versus $42$ iterations), yet at a much lower complexity. Meanwhile, setting a slightly higher iteration number for the SA-based initialization method, e.g., $I_{SA}=10$, improves the initial achievable rate and the convergence rate (i.e., $9$ iterations). Furthermore, all the proposed CPM-based methods converge much faster compared to the random phase method (i.e., $102$ iterations), while all methods finally achieve the same converged achievable rate (i.e., $1.3685$ bps/Hz). This thus validates the efficiency of the proposed CPM-based initialization methods. Hence, in the following, we adopt the SA-based method with $I_{SA}=10$ for the proposed CPM initialization and iterative method. 

\begin{figure}[t]
    \centering
    \includegraphics[width=0.6\linewidth, keepaspectratio]{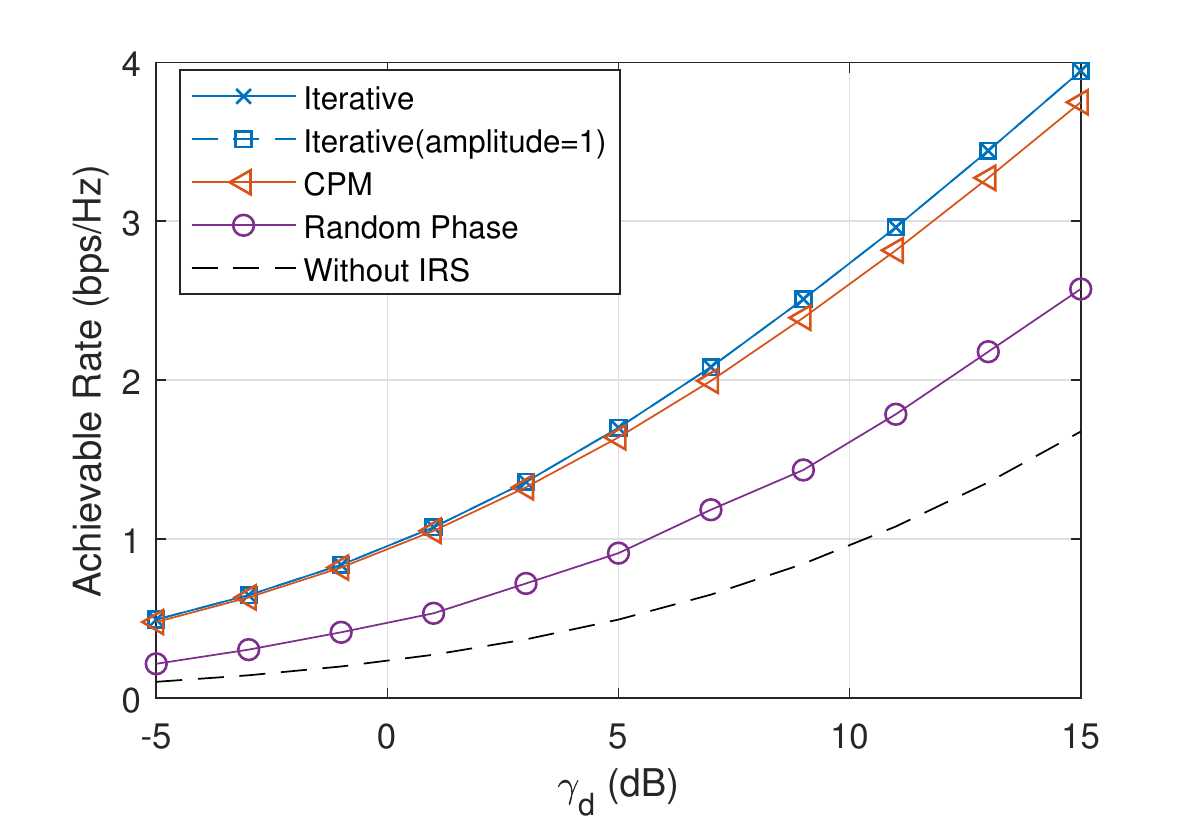}
    \vspace{-3mm}
    \caption{Achievable rate versus SNR.}
    \label{fig:su_snr}
    \vspace{-7mm}
\end{figure}
Fig. \ref{fig:su_snr} shows the performance of the iterative algorithm and the benchmark schemes at different SNR values, with $M\!=\!20$ and $\alpha\!=\!0.1$. It is observed that all the four schemes with IRS outperform the scheme without IRS, owing to the IRS-enhanced average channel power between the BS and the user. Moreover, the proposed iterative algorithm and CPM-based initialization scheme both achieve significantly improved achievable rates over the random phase scheme, since the direct channel and the reflecting channel are superposed more constructively via careful design of the IRS reflection coefficients. Meanwhile, it is observed that the reflection coefficients of the proposed alternating optimization method always have unit amplitudes at convergence, and therefore the two iterative methods have identical results. This further lowers the implementation complexity as the dedicated controller for amplitude variation of the reflection coefficients is not required and the IRS practically acts as phase shifters, while the relaxation on the amplitude simplifies the optimization design. Furthermore, it is also observed that the performance of the proposed CPM-based initialization scheme is very close to that of the iterative algorithm. Therefore, this scheme is suitable for practical implementation with lower complexity.   

\begin{figure}[t]
    \centering
    \includegraphics[width=0.6\linewidth, keepaspectratio]{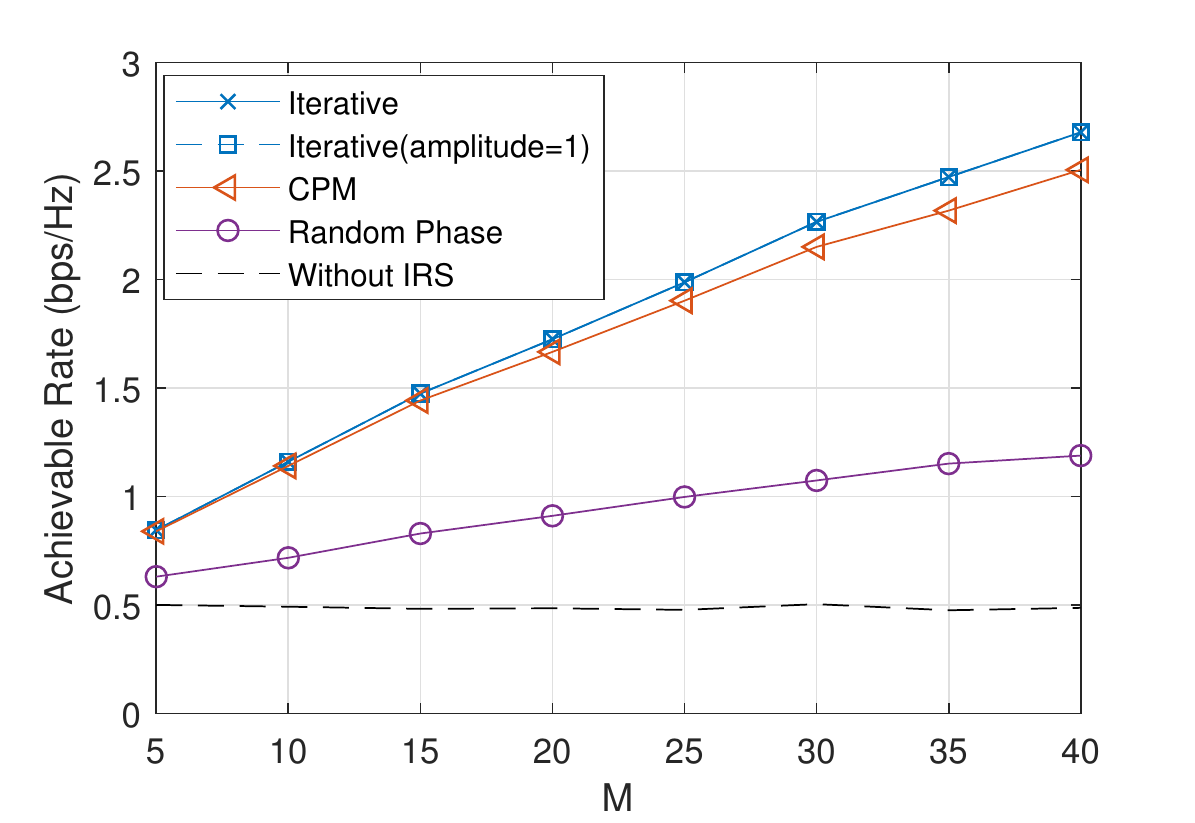}
    \vspace{-3mm}
    \caption{Achievable rate versus number of reflecting elements.}
    \label{fig:su_m}
    \vspace{-7mm}
\end{figure}

Fig. \ref{fig:su_m} compares the performance of the iterative algorithm and the benchmark schemes versus $M$, where we set $\gamma_d=5$ dB and $\alpha=0.1$. It is observed that the achievable rates for both the proposed iterative algorithm and initialization scheme increase with $M$, owing to the passive beamforming gain harvested by properly designing the IRS reflection coefficients according to the CSI; while on the other hand, the achievable rate for the random phase scheme has a slower growth rate with $M$. Moreover, it is observed that the performance gain of the proposed schemes over the scheme without IRS or with random phase becomes more pronounced as $M$ increases.

\begin{figure}[t]
    \centering
    \includegraphics[width=0.6\linewidth, keepaspectratio]{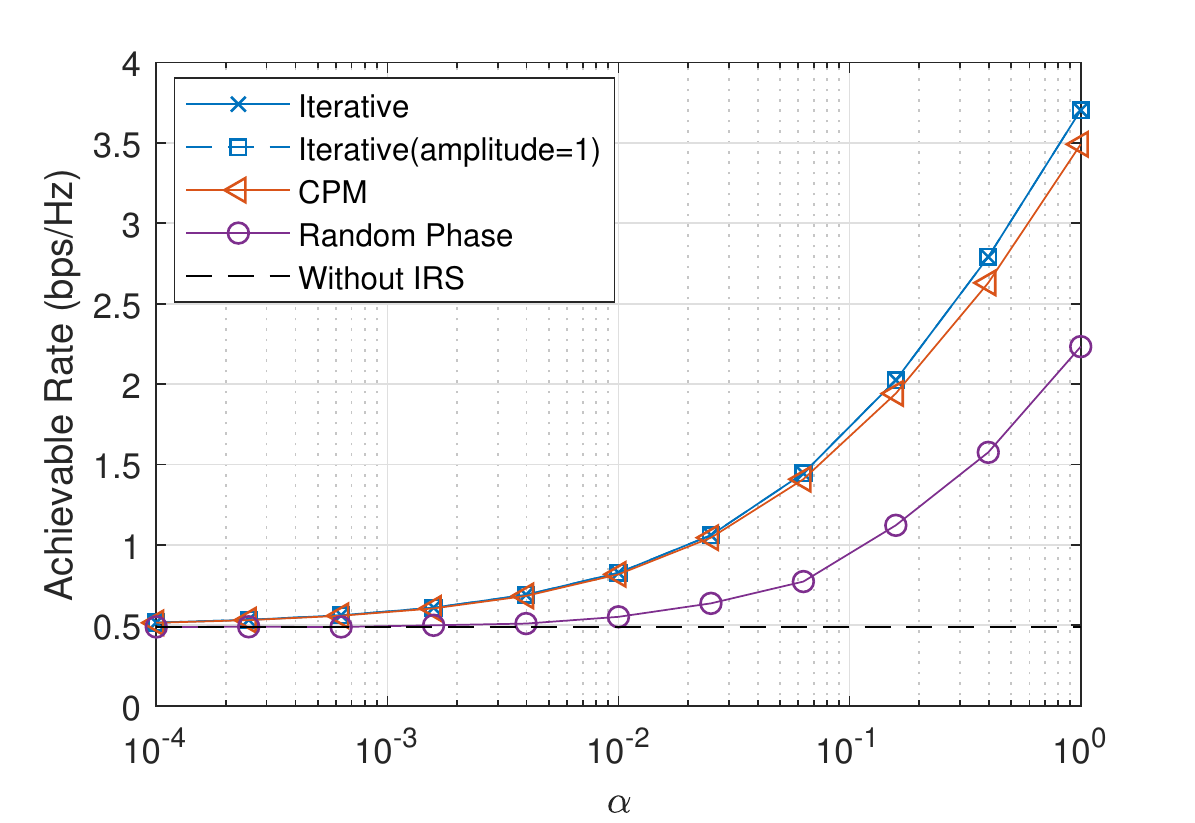}
    \vspace{-3mm}
    \caption{Achievable rate versus reflected to direct link power ratio.}
    \label{fig:su_ratio}
    \vspace{-7mm}
\end{figure}

Fig. \ref{fig:su_ratio} shows the performance of the iterative algorithm and the benchmark schemes versus the reflected to direct link power ratio $\alpha$ with $M=20$ and $\gamma_d=5$ dB. As $\alpha$ increases, this corresponds to a practical scenario where the user maintains a fixed distance with the BS (e.g., on a circle centered at the BS), and gradually moves towards the IRS. It is observed that when $\alpha$ is very small, all the schemes yield similar performance since the user is far away from the IRS, whose effect is thus negligible. In contrast, as $\alpha$ increases, the performance gains of the proposed iterative algorithm and CPM-based initialization method over the benchmark schemes increase drastically, due to the increased dominance of the IRS reflected link over the direct link. Moreover, it is observed that all the schemes with IRS achieve significantly improved performance compared to that without IRS when $\alpha$ is large. This indicates that even though the user is located far away from the BS in the cell-edge scenario, a nearby IRS is effective in enhancing the link rate.

\subsection{Performance of Proposed Algorithms with Imperfect Channel Estimation}
In this subsection, we examine the performance of the proposed protocol, and evaluate the impact of IRS grouping ratio and imperfect channel estimation to the proposed algorithms. Note that it has been shown in the previous subsection that, compared with the iterative algorithm, the proposed CPM-based initialization method incurs only marginal performance loss, yet with much lower complexity. Thus, we consider only the proposed CPM-based initialization method for practical implementation in this subsection. Also note that the plots of the achievable rate in this subsection take into account the performance loss due to training overhead and estimation error. For the purpose of drawing essential insights, we ignore the delay time due to processing and feedback (i.e., $\tau_{D}=0$), as the channel training overhead $T_p$ as well as the channel coherence time $T_c$ is generally much larger than $\tau_D$. For illustration, the number of reflecting elements is set as $M=100$ with $M_x=M_y=10$, and the reflected-to-direct link power ratio is set as $\alpha=0.1$. The grouping of IRS reflecting elements at different IRS grouping ratio is given in Table \ref{table:irs_group}. Zadoff-Chu (ZC) sequence \cite{zachu} is employed as the pilot sequence with $P_t=20P$. The other parameters are the same with the previous subsection. 

\begin{table}[t]
\centering
\caption{Grouping of IRS elements}
\begin{tabular}{ | c | c | c | c | c | c | c|}   
    \hline
    $\rho$ & $1/100$ & $1/50$ & $1/25$ & $1/10$ & $1/4$ & $1$ \\ 
    \hline
    $K$ & $1$ & $2$ & $4$ & $10$ & $25$ & $100$\\
    \hline
    $B_x\times B_y$&$10\times 10$ & $5\times 10$ & $5\times 5$ & $2\times 5$ & $2\times 2$ & $1\times 1$\\
    \hline
    \end{tabular}
\label{table:irs_group}
\end{table}

\begin{figure}[t]
	\centering
	\includegraphics[width=0.6\linewidth]{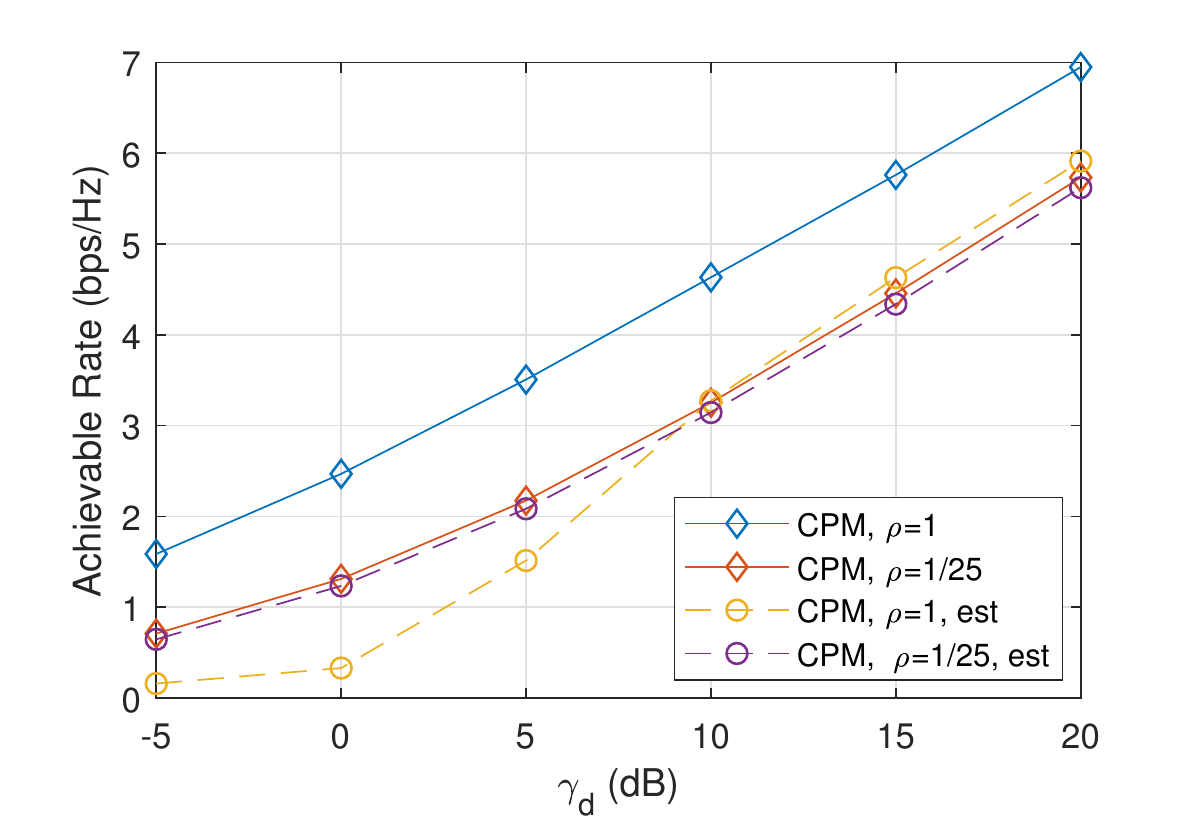}
	\vspace{-3mm}
	\caption{Achievable rate versus direct link SNR with perfect versus estimated CSI.}
	\label{fig:impact_CE}
	\vspace{-7mm}
\end{figure}
We first evaluate the impact of imperfect channel estimation on the performance of the proposed CPM-based method versus the direct link SNR $\gamma_d$ in Fig. \ref{fig:impact_CE}. Two different IRS grouping ratios are compared, with perfect CSI versus with estimated CSI. The channel coherence time is set as $T_c=900$. It is observed that the proposed CPM-based method with $\rho=1$ is more sensitive to channel estimation error, especially at low-to-medium SNR regimes. As the SNR increases, the proposed channel estimation becomes more accurate and the performance gap between the achievable rate with estimated CSI and that with perfect CSI decreases at $\rho=1$. On the other hand, it is observed that at lower grouping ratio, e.g., $\rho=1/25$, the proposed CPM-based method employing the proposed channel estimation method has a much smaller performance gap with that using perfect CSI at both high and low SNR regimes. This is because with grouping, more IRS elements are switched on each time during the channel estimation, resulting in a higher receive SNR and therefore more accurate channel estimation, as observed from \eqref{MSE}. In the following, we evaluate the performance of the proposed grouping with imperfect CSI in both high-SNR and low-SNR regimes. 

\begin{figure}[t]
	\centering
	\includegraphics[width=0.6\linewidth]{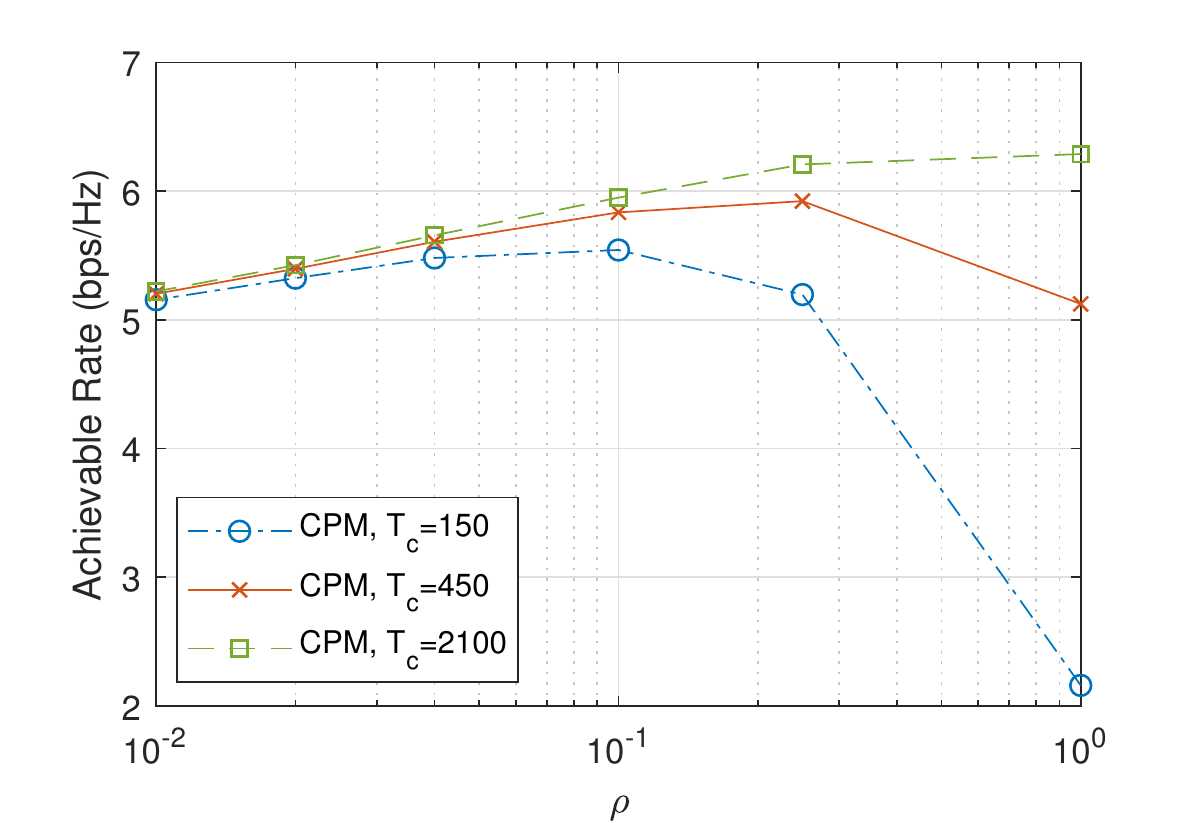}	
	\vspace{-3mm}	
	\caption{Achievable rate versus IRS grouping ratio at $\gamma_d=20$ dB (high-SNR regime).}
	\label{fig:res1}	
	\vspace{-7mm}
\end{figure}

Fig. \ref{fig:res1} evaluates the effect of grouping size (or equivalently, the IRS grouping ratio) on the achievable rate at direct link SNR $\gamma_d=20$ dB (i.e., the high-SNR regime). Note that at low grouping ratio, the number of groups $K$ is small while each group consists of a large number of IRS passive array elements sharing a common reflection coefficient, and therefore only a small overhead for channel estimation is required. As the grouping ratio increases, $K$ also increases and thus results in a larger overhead. However, due to the increased degrees of freedom for IRS coefficient design at higher grouping ratio, the reflection coefficients combine the multiple paths of the reflected link as well as the direct link more effectively, leading to a higher achievable rate in the data transmission phase. Hence, it is observed that the achievable rate in general first increases then decreases with the grouping ratio. This is because the performance improvement by higher grouping ratio is overwhelmed by the increased training overhead, especially when the latter occupies a substantial amount of time during one coherence block. In contrast, for a relatively large channel coherence time (e.g., $T_c=2100$), the training overhead constitutes an insignificant part of the coherence time and the achievable rate keeps increasing with the grouping ratio. Therefore, for practical implementation, the optimal grouping ratio critically depends on the channel coherence time.

\begin{figure}[t]
	\centering
	\includegraphics[width=0.6\linewidth]{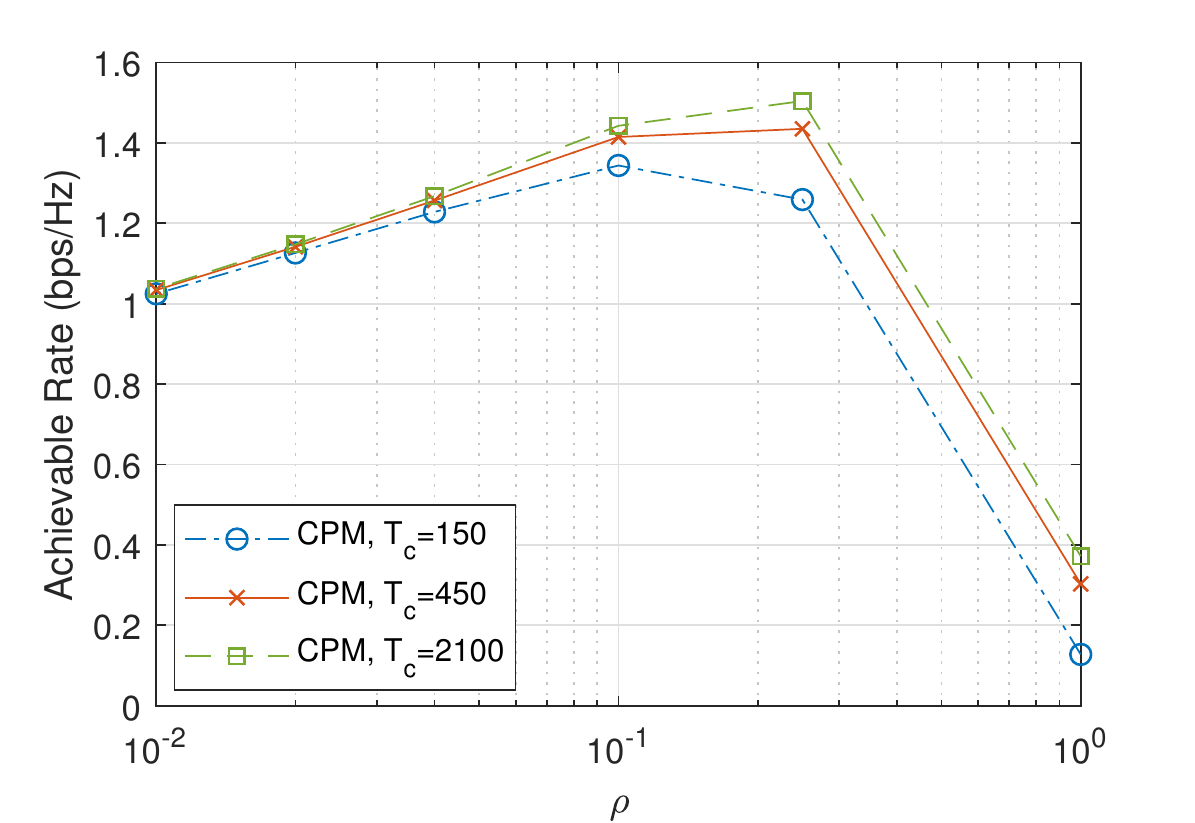}		
	\vspace{-3mm}	
	\caption{Achievable rate versus IRS grouping ratio at $\gamma_d=0$ dB (low-SNR regime).}
	\label{fig:res2}	
	\vspace{-7mm}
\end{figure}

Fig. \ref{fig:res2} shows the effect of the IRS grouping ratio on the achievable rate at direct link SNR $\gamma_d=0$ dB (i.e., low-SNR regime), where a different trend is observed compared to Fig. \ref{fig:res1} in the high-SNR regime. Specifically, for all values of the channel coherence time, the achievable rate first increases and then decreases, for which the reasons are explained as follows. First, the CSI obtained at low SNR is not accurate, and therefore leads to a greater performance loss at both low and high grouping ratios as the design variables are optimized based on the estimated CSI. Meanwhile, at low grouping ratio and low SNR, increasing the grouping ratio leads to higher diversity for the IRS beamforming and thus improved achievable rates. However, at large grouping ratio (e.g., $\rho=1$), the achievable rate gain by higher passive beamforming diversity is overwhelmed by the increased channel estimation error and overhead, resulting in a sharp rate decrease for all coherence time.

\begin{figure}[t]
	\centering
	\includegraphics[width=0.6\linewidth]{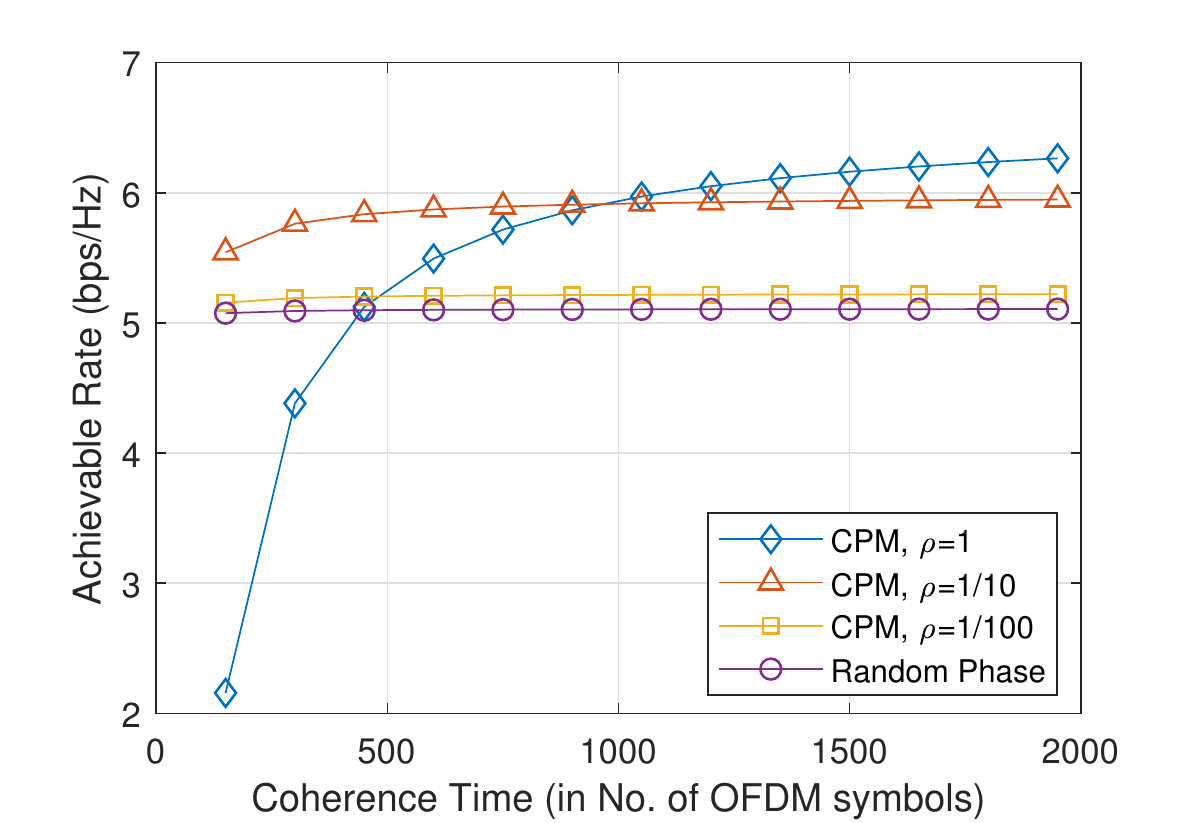}
	\vspace{-3mm}
	\caption{Achievable rate versus coherence time at $\gamma_d=20$ dB (high-SNR regime).}
	\label{fig:coh_comp1}
	\vspace{-7mm}
\end{figure}

\begin{figure}[t]
	\centering
	\includegraphics[width=0.6\linewidth]{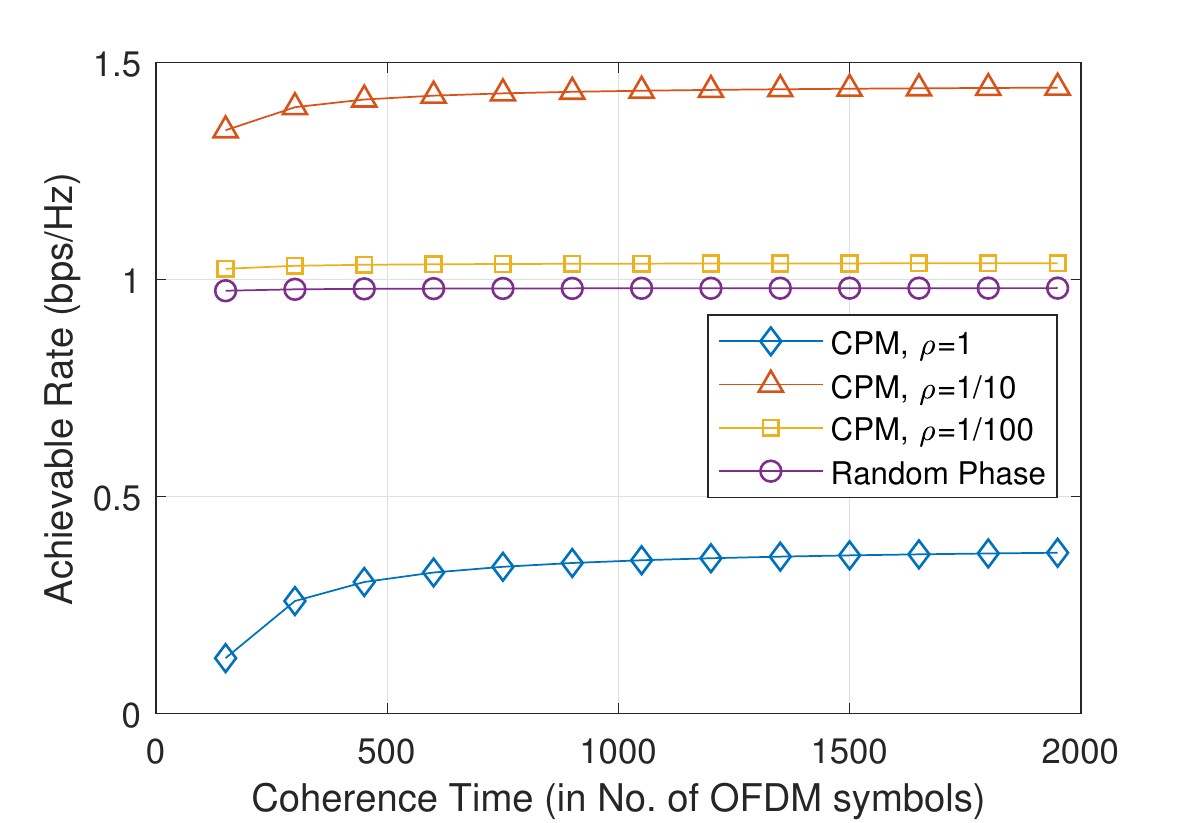}
	\vspace{-3mm}
	\caption{Achievable rate versus coherence time at $\gamma_d=0$ dB (low-SNR regime).}
	\label{fig:coh_comp2}
	\vspace{-7mm}
\end{figure}
Fig. \ref{fig:coh_comp1} further evaluates the achievable rate performance versus the channel coherence time under different grouping ratios at direct link SNR $\gamma_d=20$ dB for the proposed CPM-based method. It is observed that for the proposed protocol, the length of the coherence time is critical to the optimal IRS grouping ratio with the highest achievable rate. As the coherence time increases, the achievable rate for every IRS grouping ratio increases, while the increment for higher grouping ratio is more prominent. In other words, the optimal IRS grouping ratio increases with the coherence time, as the loss due to longer training overhead can be fully compensated by the increased effectiveness of the IRS reflection coefficients in combining the multiple paths coherently. In contrast, Fig. \ref{fig:coh_comp2} shows the achievable rate performance versus channel coherence time at direct link SNR $\gamma_d=0$ dB. Note that at such low SNR, it is more desirable to divide the IRS elements into groups, as the performance improvement by increasing the number of groups to enjoy higher beamforming gain is limited and overwhelmed by the increased overhead and channel estimation error. The benchmark scheme with random phase is also included for comparison, which only needs to estimate the combined link $\tilde{\boldsymbol{h}}$ and feed back the transmit power allocation, thus incurring a much shorter training overhead. Nevertheless, it is observed from Fig. \ref{fig:coh_comp1} and Fig. \ref{fig:coh_comp2} that the performance of the proposed algorithm with proper grouping always outperforms the random phase scheme regardless of the SNR regimes, even at extremely small coherence time.

\section{Conclusion}
\label{sec:con}
In this paper, we proposed a novel approach to enhance the achievable rate of an OFDM system by utilizing the IRS. We devised a practical transmission protocol by flexibly grouping the IRS elements and estimating the combined channel for each group, where data transmission proceeds by considering a common IRS reflection coefficient for each group. Under the proposed protocol with any given grouping, we formulated the joint optimization problem of the transmit power allocation and IRS reflection coefficients based on the estimated channels. By leveraging optimization techniques, we proposed computationally efficient methods to find high-quality suboptimal solutions for the formulated problem. Numerical results showed the effectiveness of IRS in boosting the achievable rate of a cell-edge user aided by the IRS. The proposed initialization method was also shown to achieve very close rate performance compared to the proposed iterative method, but with much lower complexity for implementation. Finally, the proposed protocol was shown to outperform the straightforward counterpart without IRS elements grouping in terms of achievable rate under various practical setups. 

There are a number of promising directions worthy of investigation in future works. For example, we consider the joint transmit power and IRS coefficients optimization with given grouping and training overhead in this work, while the grouping strategy as well as the training sequence can also be optimized to maximize the achievable rate. More advanced pilot design specifically catered to IRS-aided systems can also be investigated. Moreover, this paper considers a single-user system, while it is interesting to extend the results to more general setups such as multiuser OFDMA systems with joint multiuser resource allocation.

\bibliographystyle{IEEEtran}
\bibliography{irs}
\end{document}